\documentclass[MSNbibl,nameyear,dvips]{arxstspdf}
\usepackage{dcolumn}
\usepackage{graphicx}
\usepackage{flushend}
\usepackage{stfloats}

\volume{28}
\issue{2}
\pubyear{2013}
\firstpage{135}
\lastpage{167}
\doi{10.1214/12-STS410} 

\makeatletter
\renewcommand{\mid}{\vert}
\newcolumntype{d}[1]{D{.}{.}{#1}}

\newproclaim{defn}{Definition}[section]

\newproclaim{exa}{Example}[section]
\newproclaim{rem}{Remark}[section]

\newcommand{\BIC}{\mathit{BIC}}
\newcommand{\GIC}{\mathit{GIC}}
\newcommand{\TIC}{\mathit{TIC}}
\newcommand{\MDL}{\mathit{MDL}}
\newcommand{\PEC}{\mathit{PEC}}
\newcommand{\DIC}{\mathit{DIC}}

\newcommand{\E}{\operatorname{E}}

\newcommand{\Var}{\operatorname{Var}}
\newcommand{\trace}{\operatorname{trace}}
\newcommand{\rank}{\operatorname{rank}}
\newcommand{\diag}{\operatorname{diag}}
\renewcommand{\P}{\mathbf{P}}

\newcommand{\argmax}{\mathop{\arg\max}}

\newcommand{\bdiag}{\mathop{\mathrm{\mathrm{block} \operatorname{diag}}}}

\renewcommand{\tilde}[1]{\widetilde{#1}}
\renewcommand{\hat}[1]{\widehat{#1}}

\newcommand{\ba}{\mathbf{a}} 

\newcommand{\dd}{\mathbf{d}}
\newcommand{\ee}{\mathbf{e}}

\newcommand{\qqq}{\mathbf{q}}

\newcommand{\bt}{\mathbf{t}}
\newcommand{\uu}{\mathbf{u}}
\newcommand{\vv}{\mathbf{v}}

\newcommand{\xx}{\mathbf{x}}
\newcommand{\yy}{\mathbf{y}}
\newcommand{\zz}{\mathbf{z}}

\newcommand{\bA}{{\mathbf{A}}}
\newcommand{\BB}{{\mathbf{B}}}
\newcommand{\CC}{{\mathbf{C}}}
\newcommand{\DD}{{\mathbf{D}}}

\newcommand{\HH}{{\mathbf{H}}}
\newcommand{\II}{{\mathbf{I}}}
\newcommand{\JJ}{{\mathbf{J}}}
\newcommand{\KK}{{\mathbf{K}}}
\newcommand{\LL}{{\mathbf{L}}}

\newcommand{\PP}{{\mathbf{P}}}

\newcommand{\RR}{{\mathbf{R}}}

\newcommand{\UU}{{\mathbf{U}}}
\newcommand{\VV}{{\mathbf{V}}}

\newcommand{\XX}{{\mathbf{X}}}

\newcommand{\ZZ}{{\mathbf{Z}}}

\newcommand{\be}{\bolds{\beta}}
\newcommand{\ga}{\bolds{\gamma}}
\newcommand{\de}{\bolds{\delta}}
\newcommand{\ep}{\bolds{\varepsilon}}

\newcommand{\bth}{\bolds{\theta}} 

\newcommand{\ta}{\bolds{\tau}}

\newcommand{\ph}{\bolds{\phi}}

\newcommand{\ps}{\bolds{\psi}}

\newcommand{\GA}{\bolds{\Gamma}}
\newcommand{\DE}{\bolds{\Delta}}
\newcommand{\bTH}{\bolds{\Theta}} 

\newcommand{\SI}{\bolds{\Sigma}}

\newcommand{\PH}{\bolds{\Phi}}
\newcommand{\PS}{\bolds{\Psi}}

\newcommand{\one}{\mathbf{1}}
\newcommand{\zer}{\mathbf{0}}
\makeatother

\begin{document}
\begin{frontmatter}

\title{Model Selection in Linear Mixed Models}
\runtitle{Model Selection in Linear Mixed Models}

\begin{aug}
\author[a]{\fnms{Samuel} \snm{M\"uller}\ead[label=e1]{samuel.mueller@sydney.edu.au}},
\author[b]{\fnms{J. L.} \snm{Scealy}\ead[label=e2]{Janice.Scealy@anu.edu.au}}
\and
\author[c]{\fnms{A. H.} \snm{Welsh}\corref{}\ead[label=e3]{Alan.Welsh@anu.edu.au}}
\runauthor{S. M\"uller, J.~L. Scealy and A. H. Welsh}

\affiliation{University of Sydney, Australian National
University and Australian National
University}

\address[a]{Samuel M\"uller is Senior Lecturer, School of Mathematics
and Statistics F07,
University of Sydney, NSW 2006, Australia \printead{e1}.}
\address[b]{J.~L. Scealy is Postdoctoral Research Associate, Centre for
Mathematics and its Applications, Australian National
University, Canberra, ACT 0200, Australia \printead{e2}.}
\address[c]{A.~H. Welsh is E.~J. Hannan Professor of Statistics,
Centre for Mathematics and its Applications, Australian National
University, Canberra, ACT 0200, Australia \printead{e3}.}
\end{aug}

%
\begin{abstract}
Linear mixed effects models are highly flexible in handling a broad
range of data types and are therefore widely used in applications. A
key part in the analysis of data is model selection, which often aims
to choose a parsimonious model with other desirable properties from a
possibly very large set of candidate statistical models. Over the last
5--10 years the literature on model selection in linear mixed models has
grown extremely rapidly. The problem is much more complicated than in
linear regression because selection on the covariance structure is not
straightforward due to computational issues and boundary problems
arising from positive semidefinite constraints on covariance matrices.
To obtain a better understanding of the available methods, their
properties and the relationships between them, we review a large body
of literature on linear mixed model selection. We arrange, implement,
discuss and compare model selection methods based on four major
approaches: information criteria such as AIC or BIC, shrinkage methods
based on penalized loss functions such as LASSO, the Fence procedure
and Bayesian techniques.
\end{abstract}

%
\begin{keyword}
\kwd{AIC}
\kwd{Bayes factor}
\kwd{BIC}
\kwd{Cholesky decomposition}
\kwd{fence}
\kwd{information criteria}
\kwd{LASSO}
\kwd{linear mixed model}
\kwd{model selection}
\kwd{shrinkage methods}
\end{keyword}\vspace*{-3pt}

\end{frontmatter}

\section{Introduction} \label{secIntroduction}\vspace*{7pt}

The class of linear mixed models (\citeauthor{Henderson1950}\break (\citeyear{Henderson1950})) provides a flexible
framework for modeling a wide range of data types, including clustered,
longitudinal and spatial data. This framework is increasingly widely
used in Applied Statistics. It is interesting and important both in its
own right and as a starting point for the development of more
complicated classes of models such as generalized linear mixed models
or GLMMs (e.g., \cite*{McCulloch2003}), nonlinear mixed models (e.g.,
\cite*{Pinheiro2000}), and various semi-parametric and
nonparametric models (e.g., Ruppert, Wand and Carroll, \citeyear{Ruppert22003}). In practical
applications of statistical models (including linear mixed models), a
key aspect of the analysis is often model selection, the choice of a
particular model within a class of candidate models; see \citet{Claeskens2008}
for a general review. With the increasing use of linear
mixed models in practice, the need to do model selection has resulted
in the implementation of a number of different methods for model
selection in software packages (such as R or SAS). There are, however,
other, recent methods which have not yet been implemented in standard
software and there is no consensus in the statistical community on how
to approach model selection for linear mixed models. This makes it very
difficult for an analyst to answer the basic question: Which methods
should I use and when should I use them? In this paper, as a step
toward addressing these issues, we review, classify and compare a
number of methods for selecting linear mixed models so that we can
better understand their properties and the relationships between them.

There is a substantial literature on model selection for linear mixed
models which has grown extremely rapidly in the last 5--10 years. As a
consequence of this rapid growth, researchers working in parallel in
the area have not had access to the developments of other researchers.
The inevitable result is a lack of cross-referencing between papers
using different methods for model selection, between papers using
similar methods for model selection and even between papers using
similar methods written by the same author. The main consequences are a
limited acknowledgement of other relevant work, a limited exploration
of the relationships between different methods and limited comparisons
between different methods of model selection, either theoretically or
through simulation. In addition, papers treating the same model use
different notation and terminology; papers proposing different
approaches do so for different models (including special cases of
general models obtained either by imposing special structure or by
treating some parameters as known) or treat different types of
selection problems (such as only selecting the regression parameters),
making it difficult to access and evaluate the key methods.
Finally, only a few papers discuss and solve computational issues. We
do not give specific examples here because our intention is not to
single out any particular contributions but rather to describe the
state of the literature as a whole.

Linear mixed models can be viewed as extensions of linear regression
models, so
many of the methods proposed for selecting mixed models can be seen as
extensions of methods developed for linear regression models. However,
this does not mean that model selection for linear mixed models can be
subsumed within model selection for linear regression models. It is
useful to exploit the similarities between the models but there are
also important differences between linear mixed models and linear
regression models which need to be taken into account. In linear
regression models, the responses are independent, whereas, in linear
mixed models, they are typically dependent. This dependence impacts on
model selection by reducing the effective sample size, a quantity that
affects the theoretical properties of procedures and is used explicitly
in some model selection procedures such as the Bayesian Information
Criteria (BIC; \cite*{Schwarz1978}) described in Section \ref
{secinformation}. The dependence also means that linear mixed models
have both regression parameters (which describe the mean structure) and
variance parameters (which describe the sources of variability and the
dependence structure). If, as is often the case, these parameters have
a different relative importance in the analysis, this should be
reflected in model selection. For example, if we are evaluating a model
for its predictive ability, it may be less important to get the
dependence structure exactly correct than it is to get the regression
structure correct. Even if we do not explicitly assign different
relative importance to the parameters, it is already implicit in the
model---it underlies the familiar difficulty of assigning degrees of
freedom or measuring model complexity in linear mixed models. It is
also often the case that regression parameters are unconstrained,
whereas variance parameters are always constrained by the requirement
that variance matrices must be positive semi-definite. In many
problems, many of the parameters are required to be nonnegative so
there are boundaries of the parameter space at zero. An important part
of model selection is setting a parameter to zero which, unfortunately,
means putting some of the variance parameters on the boundary.
Consequently, there are boundary issues in model selection with
variance parameters, either computational issues from fitting models
with redundant variance parameters (as software tends not to handle
this well) or statistical issues related to testing null hypotheses on
the boundary of the parameter space (because selection is closely
related to hypothesis testing; \cite*{Claeskens2008}), that do not
arise when selecting regression parameters. Thus, model selection for
linear mixed models is different from model selection for linear
regression models and it is important to acknowledge and take into
account the differences between the two classes of models.

For the linear regression model there is a large and growing literature
on variable selection in the high-dimensional setting (e.g., \cite*{Fan2010};
B\"uhlmann and van de Geer, \citeyear{Buhlmann2011}). This is very different
from the fixed (finite) dimensional case because many of the fixed
dimensional model selection procedures either do not work at all or,
for their implementation, require some theoretical or computational
adjustment. Additional assumptions such as sparsity in the true model
are also needed in the high-dimensional\vadjust{\goodbreak} setting in order to obtain
consistent model selection.
Nonetheless, sometimes similar methods can be used in both the high and
fixed-dimensional cases, for example, shrinkage methods based on the
LASSO (\cite{Tibshirani1996}) are used extensively in both contexts.
To date, most of the literature on model selection for the linear mixed
model is for the fixed-dimensional parameter case and it is only very
recently that authors have started to consider high-dimensional
settings (Schelldorfer, B{\"u}hlmann and van~de Geer, \citeyear{Schell2011}; \cite*{Fan2012}). Part of the
reason for this lack of coverage is because asymptotic studies in the
high-dimensional linear mixed model case are more difficult than in the
linear regression case since both the number of regression parameters
and/or variance parameters can potentially grow with the sample size
and at possibly different rates. There are also more complex
computational and estimation issues to consider due to the presence of
large, sparse covariance matrices.

In this paper we review model selection for linear mixed models
focusing mostly on the fixed-dimen\-sional parameter case. We define
these models formally, distinguish different model selection problems
for the models and introduce the basic notation in Section \ref
{secnotation}. We classify the different methods into four broad
approaches and describe each approach in its own section. The first
approach is based on choosing models to minimize information criteria
such as the widely used Akaike Information Criteria (AIC; \cite*{Akaike1973})
and the Bayesian Information Criteria (BIC; \cite*{Schwarz1978}). These
criteria are described in Section~\ref{secinformation}. We describe
shrinkage methods like the LASSO (\cite{Tibshirani1996}) in Section \ref
{seclasso} and the Fence method (\cite{Jiang2008}) in Section \ref
{secfence}. We briefly discuss some Bayesian methods in Section~\ref
{secBayes}. Finally, we review some published simulation results in
Section~\ref{secsimulation} and conclude with discussion and
conclusions in Section~\ref{secdiscussion}.

Although model selection can be formulated and interpreted in terms of
testing, we do not review testing per se in this paper. There is a huge
literature on testing, a substantial part of which could be construed
to have at least some relevance to model selection, and we simply have
to draw a line somewhere. We therefore focus on methods which may be
motivated by and derived from tests but ultimately do not explicitly
focus on tests. Second, our focus is on the ideas behind and the
relationships between methods, rather than the details of the
implementation of any particular method. We do identify areas of
difficulty where more work is needed, including numerical and
implementation issues, but these are not our main focus, and resolving
them in this paper is even further from our main focus. In particular,
any discussion of Bayesian methods leads quickly toward computation,
but we do not review Bayesian computation.

\section{The Model Selection Problem} \label{secnotation}

Consider the linear mixed model
%
\begin{equation}
\yy= \XX\be+ \ZZ\GA\uu+ \DE\ee, \label{eqremodel}
\end{equation}
where $\yy$ is a $n$-vector of observed responses, $\XX$ is a known $n
\times p$ matrix of covariates, $\ZZ$ is a known $n \times s$ matrix,
$\uu$ and $\ee$ are unobserved independent $s$ and $n$-vectors of
independent random variables with mean zero and variance the identity
matrix, $\be$ is a $p$-vector of unknown regression parameters, $\GA$
is an $s\times s$ matrix which contains $q_{\gamma}$ distinct unknown
parameters and $\DE$ is an $n \times n$ matrix which contains $q_{\delta
}$ distinct unknown parameters. Writing the model this way is motivated
by \citet{Chen2003}, Field, Pang and Welsh (\citeyear{Field2010}), Bondell, Krishna and
Ghosh (\citeyear{Bondell2010})
and \citet{Ibrahim2011}. Let $\PS= \GA\GA^{T}$ and $\SI= \DE\DE^T$
so we can write
\[
\E(\yy) = \XX\be \quad\mbox{and}\quad \Var(\yy) = \VV= \ZZ\PS\ZZ^T + \SI.
\]
The notation is general enough to allow the matrix square roots $\GA$
and $\DE$ to be the symmetric matrices produced by taking the square
roots of the eigenvalues in the spectral decomposition of $\PS$ or $\SI
$, the lower triangular matrices produced by the Cholesky decomposition
of $\PS$ or $\SI$, or, if $\PS$ is block diagonal, the block diagonal
matrix of the lower triangular matrices from the Cholesky
decompositions of each block. It is simpler to specify and interpret
the model in terms of $\PS$ and $\SI$, but it is simpler to fit and
select models with $\GA$ and $\DE$. Let $\ga$ denote the $q_{\gamma}$
distinct unknown parameters in $\GA$ and $\de$ the $q_{\delta}$
distinct unknown parameters in $\DE$. It is sometimes convenient to
group the parameters into the vector of regression parameters $\be$,
the vector of variance parameters $\ta= (\ga^T, \de^T)^T$ of length $q
= q_{\gamma} + q_{\delta}$ and the vector of all parameters $\bth= (\be^T, \ta^T)^T$ of length $p+q$.

There are other useful parametrizations for (\ref{eqremodel}) which
are used in the literature. One of these involves writing $\GA$ as
%
\begin{equation}
\label{eqmodchol} \GA=\DD\GA^{\dagger},
\end{equation}
where $\GA^{\dagger}$ is lower triangular with ones on the diagonal and
$\DD$ is a diagonal matrix (\cite{Chen2003}).\vadjust{\goodbreak} When $\SI=\sigma^2\II_n$ with $\II_n$ the $n\times n$ identity matrix, it is sometimes
convenient to write $\GA=\sigma\DD^{\dagger}\GA^{\dagger}$, where $\DD^{\dagger} = \DD/\sigma$
(Bondell, Krishna and\break Ghosh, \citeyear{Bondell2010};
Saville, Herring and Kaufman, \citeyear{Saville11}).
To be consistent with the terminology of \citet{Pou2011}, we will
refer to these as alternative Cholesky factors. The main advantage of
the alternative Cholesky parametrization is that it separates and
therefore encourages different treatment of the diagonal and the
off-diagonal elements of $\GA$. In particular, a zero diagonal element
makes the whole row zero, whereas a zero off-diagonal element affects
only itself. However, it is important to keep in mind that the diagonal
elements of $\GA$ include off-diagonal elements of $\PS$ so the order
of rows and columns in $\PS$ can affect model selection.

An alternative to the linear mixed model (\ref{eqremodel}), which is
widely used in the econometric literature, can be written as
%
\begin{equation}
\yy= \XX\be+ \VV^{1/2}\ep, \label{eqtrmodel}
\end{equation}
where $\ep$ is an $n$-vector of independent random variables with mean
zero and variance one. Models (\ref{eqremodel}) and~(\ref{eqtrmodel})
have the same mean and variance. If all the random variables ($\uu$,
$\ee$, $\ep$) have Gaussian distributions, the responses $\yy$ in
models (\ref{eqremodel}) and (\ref{eqtrmodel}) have the same
distribution. However, the two models are not necessarily identical
because they can have different parameter spaces; the parameter space
for (\ref{eqremodel}) requires $\PS$ to be positive definite, whereas
that for (\ref{eqtrmodel}) only requires $\VV$ to be positive
definite. Thus, the parameter space for~(\ref{eqtrmodel}) can be
larger than and contain that for (\ref{eqremodel}). If any of the
random variables have non-Gaussian distributions, then the responses in
the two models have the same first two moments but can have different
higher order moments and different distributions (\cite{Field2007}), as well as different parameter spaces. We call (\ref
{eqtrmodel}) the transformation model to be consistent with \citet{Field2007};
it is sometimes called the marginal model (e.g., \cite*{jiang2007}). The difference between the two models is not widely appreciated,
but it is important to be clear about which model each procedure is
working with. Most model selection procedures have been derived for the
linear mixed model~(\ref{eqremodel}), but some of them also apply to
the transformation model~(\ref{eqtrmodel}).

It is useful to identify some special cases of the model because these
give insight into the range of forms of the model and because we will
refer to them specifically in what follows. We express these as special
cases of the linear mixed model (\ref{eqremodel}); they can also be
expressed as special cases of the transformation model (\ref{eqtrmodel}).

\textit{Variance component model} (\cite{Henderson1950}): $\PS= \bdiag
(\gamma_1^2\II_{r_1}, \ldots, \gamma_{q_{\gamma}}^2 \II_{r_{q_{\gamma
}}})$, where $s=\break \sum_{k=1}^{q_{\gamma}} r_k$. Write $\ZZ= (\ZZ^{(1)},
\ldots, \ZZ^{(q_{\gamma})})$, where $\ZZ^{(k)}$ is $n \times r_k$, and
$\uu= (\uu_1^T, \ldots, \uu_{q_{\gamma}}^T)^T$, where $\uu_k$ is a
$r_k$-vector, so that 
%
\begin{equation}\quad
\yy= \XX\be+ \gamma_1\ZZ^{(1)} \uu_1 +
\cdots+ \gamma_{q_{\gamma}}\ZZ^{(q_{\gamma})}\uu_{q_{\gamma}} + \DE\ee.
\label{eqremodelvc}
\end{equation}
Often, $\SI$ is known up to an unknown constant; in this case $q_{\delta
}=1$ and we can write $\SI= \RR_0 + \delta^2\RR_1$, with $\RR_0$ and
$\RR_1$ known. It is most common to have $\RR_0=\zer$ and $\RR_1=\II_n$, the $n \times n$ identity matrix, but other possibilities do
occur. The parameters $\gamma_1^2, \ldots,\break \gamma_{q_{\gamma}}^2, \delta^2$\vspace*{1pt} are known as variance components.

\textit{Independent cluster model}: $\PS= \bdiag(\PS_{1}, \ldots,\break \PS_{m})$, where $\PS_i$ is $s_i \times s_i$ and $s = \sum_{i=1}^{m}
s_i$, and $\SI= \bdiag(\SI_{1}, \ldots, \SI_{m})$,\vspace*{1pt} where $\SI_i$ is
$n_i \times n_i$ and $n = \sum_{i=1}^{m} n_i$. Write $\yy= (\yy_1^T,
\ldots, \yy_{m}^T)^T$,\vspace*{1pt} where $\yy_i$ is an $n_i$-vector,
$\XX= (\XX_1^T, \ldots, \XX_{m}^T)^T$,\vspace*{1pt} where $\XX_i$ is an $n_i \times p$ matrix,
$\ZZ= \bdiag(\ZZ_1, \ldots, \ZZ_{m})$, where $\ZZ_i$ is an $n_i
\times s_i$ matrix, and $\uu= (\uu_1^T, \ldots, \uu_{m}^T)^T$,\vspace*{1pt} where
$\uu_i$ is an $s_i$-vector, and $\ee= (\ee_1^T, \ldots, \ee_{m}^T)^T$,
where $\ee_i$ is an $n_i$-vector. Then, if $\GA$ and $\DE$ are block
diagonal square roots of $\PS$ and $\SI$ with $\GA_i$ and $\DE_i$ on
the diagonal, respectively, we can write (\ref{eqremodel}) as
%
\begin{equation}
\yy_i = \XX_i\be+ \ZZ_i\GA_i
\uu_i + \DE_i\ee_i,\quad  i = 1,\ldots , m.
\label{eqremodelg}
\end{equation}
The observations $\yy_1,\ldots, \yy_m$ from distinct clusters are
independent random vectors.

The independent cluster model is also called the Laird--Ware model,
though perhaps this should be restricted to the case with constant
$s_i$ (\cite{Laird1982}). The assumption of independence between
clusters makes the model easier to work with than spatial and other
models with more complete dependence structures. For this reason, much
of the work on linear mixed models and model selection for linear mixed
models has been carried out for the independent cluster model.

\textit{Clustered variance component model}: A combination of the
variance component model and the independent cluster model obtained as
a special case of the independent cluster model with $\PS_i = \break\bdiag
(\gamma_1^2\II_{r_{i1}},  \ldots, \gamma_{q_{\gamma}}^2 \II_{r_{iq_{\gamma}}})$,\vspace*{1pt} where
$s_i= \sum_{k=1}^{q_{\gamma}} r_{ik}$.
Write $\ZZ_i = (\ZZ_i^{(1)},  \ldots, \ZZ_i^{(q_{\gamma})})$,\vspace*{1pt} where
 $\ZZ_i^{(k)}$ is $n_i \times r_{ik}$, and $\uu_i = (\uu_{i1}^T, \ldots,  \uu_{iq_{\gamma}}^T)^T$,\vspace*{1pt} where $\uu_{ik}$ is a $r_{ik}$-vector. Then we
can write (\ref{eqremodel}) as
%
\begin{eqnarray} \label{eqremodelgvc}
\yy_i &=& \XX_i\be+\gamma_1
\ZZ_i^{(1)} \uu_{i1} + \cdots
\nonumber
\\[-8pt]
\\[-8pt]
\nonumber
&&{}+
\gamma_{q_{\gamma}}\ZZ_i^{(q_{\gamma})}\uu_{iq_{\gamma}} +
\DE_i\ee_i,\quad  i = 1,\ldots, m.
\end{eqnarray}

\textit{Random intercept and slope regression model}:\break A~special
case of the clustered variance component model where the first column
of $\XX_i$ is $\one_{n_i}\!=\!(1,\ldots,1)^T$ and the $\ZZ_i^{(k)}$ are
equal to the columns of $\XX_i$. It has $s=mp$ and $q_{\gamma}=p$. We
also include the model in which the $\ZZ_i^{(k)}$ include the column of
ones and a (nonempty) subset of the columns of $\XX_i$. We call the
model with $\ZZ_i = \ZZ_i^{(1)} = \one_{n_i}$ the random intercept
regression model; it is also sometimes called the nested error
regression model. It has $s=m$ and $q_{\gamma}=1$. In the multilevel
model literature (e.g., \cite*{Snijders1999}), it is common to
allow the random intercept and slopes to be correlated, but they are
usually treated as independent in the general literature.

\textit{Fay--Herriot model} (\cite{Fay1979}):\break A~special
case of the random intercept regression model with $n_i=1$, $\ZZ= \II_n$, $\PS= \gamma^2\II_n$
and $\SI= \break\diag(r_{1},\ldots, r_n)$ is
known, so $q=q_{\gamma}=1$. Here $s=n$ and the matrix $\SI$ is assumed
known because it is not identifiable.

\textit{Longitudinal autoregression model}: A special case of the
independent cluster model with $s_i = n_i$, $\ZZ_i = \II_{n_i}$ and $\PS_i=(\psi_{ijk})$ is the $n_i \times n_i$ matrix where
\[
\psi_{ijk} = %
\cases{ \sigma^2, & $\mbox{$j =k$},$
\vspace*{2pt}
\cr
\sigma^2 \phi^{|j-k|}, & $\mbox{$j \neq k,$}$ }
\]
with $-1 < \phi< 1$, $ 1\le j,k \le n_i$.
Thus, $\ga= (\sigma, \phi)$ and $q_{\gamma} = 2$, $q_{\delta}=1$.

\textit{Linear regression model}: A special case of all the above
models but a trivial linear mixed model, the linear regression model
has $\GA= \zer$ and $\DE= \sigma^2\II_n$.

We consider the selection of linear mixed models $M\in\mathcal{M}$,
where $\mathcal{M} = \{M_l\dvtx l > 1\}$ is a countable set of distinct
models which we call candidate models. Unlike in regression models, we
cannot uniquely identify a model $M$ by its nonzero parameter vector
$\bth_M = (\be^T_M, \ta^T_M)^T$,
because setting one element of $\ta_M$ equal to zero may allow other
(redundant) elements to take arbitrary values. For example, in the
longitudinal autoregression model, if $\sigma^2=0$, then the parameter
$\phi$ is arbitrary, although any choice of $\phi$ gives the same
model. We adopt the convention of setting redundant parameters equal to
a convenient, problem specific value (such as zero if it is part of the
parameter space) so we can still distinguish models by their nonzero parameters.
Some parameters are naturally grouped together (such as the
coefficients for different levels of a factor) and it is useful in
model selection to treat them as a group rather than as separate
parameters. Also, some of the parameters such as the intercept,\vadjust{\goodbreak}
coefficients of particular variables, the error variance $\sigma^2$
when $\SI=\sigma^2\II_n$ or specific covariance parameters can be
retained in all models $M\in\mathcal{M}$. An extreme version of this
occurs when the variance structure can be regarded as known from the
way the data are collected (e.g., from the structure of the
experiment), so is held fixed in $\mathcal{M}$. (It is generally less
meaningful to select across the variance structure while retaining all
the regression parameters in the model.) We will take it as understood
that, depending on the context, the definition of $\mathcal{M}$
encompasses a range of possibilities. When a data generating model
$M_t$ exists we call it the true model and any model $M_l$ that is more
complex than the true model and satisfies $M_t \subseteq M_l$ (or $\bth_{M_t} \subseteq\bth_{M_l}$)
is called a correct model. We denote the
set of correct models $\mathcal{M}_c$. We assume that the complexity
(sometimes called the dimensionality or cardinality) of a model, $d_M$,
can be calculated and satisfies $d_{M_1} < d_{M_2}$ if $M_1\subset
M_2$. We will show later (see Section~\ref{condAIC}) that model
complexity depends on the data, the model and sometimes on the
estimation or model selection technique. It can be useful to identify a
fixed (or full) model $M_f$, which has maximal model complexity and can
be used as the initial model in stepwise model selection algorithms or
to calculate initial parameter estimates, for example, for the
Adaptive LASSO (Section~\ref{seclasso}).

We have described the model selection problem in terms of the set
$\mathcal{M}$ or in terms of the parameters of the models in $\mathcal
{M}$. The problem can also be described in terms of variables and,
while these are similar, it turns out that they are not necessarily the
same. When we describe the problem in terms of selecting variables
rather than parameters, we focus on selecting columns or groups of
columns in $\XX$ and/or $\ZZ$. Selecting columns of $\XX$ is the same
as selecting nonzero parameters in $\be$, but selecting columns of $\ZZ
$ is the same as selecting whole rows of $\GA$ (and hence rows and
columns of~$\PS$) rather than selecting individual nonzero parameters
in~$\GA$. This is shown neatly by our writing the relevant term in the
model as $\ZZ\GA\uu$ and highlights one of the important differences
between the regression and the variance parameters (which makes model
selection in linear mixed or transformation models different from model
selection in linear regression models). In terms of the alternative
Cholesky factors, selecting columns of $\ZZ$ is equivalent to selecting
the diagonal elements of $\DD$ or $\DD^{\dagger}$ while treating the
terms in $\GA^{\dagger}$ as nuisance parameters. Selection on $\be$ or
$\XX$ is sometimes called selecting fixed effects, while selection on\vadjust{\goodbreak}
$\ZZ$ is sometimes called selecting random effects. This is slightly
misleading terminology because we are not directly selecting components
of the random effects $\uu$ and it is not really applicable to the
transformation model (\ref{eqtrmodel}) which does not include random
effects. We will consider the more general problem of selecting on the
parameters $\bth$ and refer to selecting regression parameters $\be$
and variance parameters $\ta$ rather than to selecting fixed or random effects.

Model selection is often carried out by choosing models in $\mathcal
{M}$ that minimize a specific criterion. This usually involves a
trade-off between the closeness of the fit to the data and the
complexity of the model. As a practical matter, since the ultimate use
of a selected model may be different from that for which it is
selected, it may be useful to consider several criteria (as was done
explicitly for the linear regression model in M\"uller and Welsh, \citeyear{Mueller2010})
and in fact include other considerations such as the performance in
diagnostic plots.

The important problem of specifying the distributions of the random
variables in a model is not usually regarded as part of model
selection. Insofar as model selection is both a selection of the model
and the method of estimation being used to fit the model, it can
implicitly also involve a choice of underlying distributions, although
it would be better if this choice were taken seriously and made more
explicit, as it should also affect the choice of model selection
method. Most of the papers on model selection of linear mixed models
assume that all the distributions are Gaussian, although some do
explore the effect of non-Gaussian distributions in simulations (e.g.,
Dimova, Markatou and
Talal, \citeyear{Dimova2011}; \cite*{Kubokawa2011}). One exception is Ahn, Zhang and Lu
(\citeyear{Ahn2011}) who propose a model selection method based on moment estimation
which does not require any distributional assumptions.

In addition to thinking about how we want to select a model, we also
need to think about how we evaluate model selection methods. If we use
the criterion which defines one of the model selection methods, then we
bias the evaluation in favor of that method. This is noted by M\"uller
and Welsh (\citeyear{Mueller2005,Mueller2009}) in the context of robust model selection. For
this reason, we suggest using criteria which are not directly related
to the definition of any specific method. These include the probability
of selecting the true model, the probability of selecting a model from
a subset of correct models in the neighborhood of the true model, the
probability of selecting a correct model (Jiang, Nguyen and Rao, \citeyear{Jiang2008,Jiang2009}) or
the mean squared error of the difference between the predictions from
the selected model and the predictions from the true model fitted by
maximum likelihood estimation (Bondell, Krishna and
Ghosh, \citeyear{Bondell2010}; \cite*{Ibrahim2011}). The performance of the model selection methods usually depends
on the class of candidate models $\mathcal{M}$, the true model and the
data. As with linear regression models, no single method for model
selection will always perform best.

For the linear mixed model (\ref{eqremodel}), the log density of $\yy$
given $\uu$ viewed as a function of the parameters is sometimes called
the conditional log-likelihood. If $\ee$ has a Gaussian distribution,
the conditional log-likelihood is
%
\begin{eqnarray}\label{condlogliktheta}
\ell(\bth| \uu) &=& -\tfrac{1}{2} \bigl\{\log|\SI| + (\yy- \XX\be- \ZZ \GA
\uu)^T
\nonumber
\\[-8pt]
\\[-8pt]
\nonumber
&&\hspace*{42pt}{}\cdot\SI^{-1}(\yy- \XX\be- \ZZ\GA\uu) \bigr\}
\end{eqnarray}
and, for simplicity, we omit here and below the constant $-\frac
{n}{2}\log2\pi$ term. Let $\langle\uu\rangle$ denote the density of~$\uu$.
If $\uu$ has a Gaussian distribution, the log-likeli\-hood
(sometimes called the marginal log-likelihood) is
%
\begin{eqnarray}\label{logliktheta}
\qquad\ell(\bth) &=& \log \biggl[\int\exp\bigl\{\ell(\bth| \uu) \bigr\}\langle\uu
\rangle \,d\uu \biggr]
\nonumber
\\[-8pt]
\\[-8pt]
\nonumber
&=& -\frac{1}{2} \bigl\{\log|\VV| + (\yy- \XX
\be)^T \VV^{-1}(\yy- \XX\be) \bigr\}.
\end{eqnarray}
This is also the log-likelihood of the Gaussian transformation model
(\ref{eqtrmodel}). For fixed $\ta$, the log-likelihood $\ell(\bth)$ is
maximized over $\be$ by the generalized least squares estimator
%
\begin{equation}
\label{eqgls} \hat{\be}(\ta) = \bigl(\XX^T\VV^{-1}\XX
\bigr)^{-1}\XX^T\VV^{-1}\yy.
\end{equation}
Modifying the profile log-likelihood $\ell(\hat{\be}(\ta), \ta)$ by
including a bias adjustment yields the useful restricted maximum
likelihood (REML) criterion function
\begin{eqnarray*}
\ell_R(\ta) = -\tfrac{1}{2} \bigl\{ \log|\VV| + \log|
\XX^T\VV^{-1}\XX| + \yy^T \PP^{-1}\yy
\bigr\},
\end{eqnarray*}
where $\PP= \VV^{-1} - \VV^{-1}\XX(\XX^T\VV^{-1}\XX)^{-1}\XX^T\VV^{-1}$ (\cite{Patterson1971}).
Let $\hat{\be}$ and $\hat{\ta}$ be maximum likelihood estimators of $\be
$ and $\ta$, respectively, and let $\hat{\ta}_R$ be a REML estimator of
$\ta$. Put
$\hat{\be}_R = \hat{\be}(\hat{\ta}_R)$.

Many of the desirable properties of maximum likelihood and REML
estimators are asymptotic properties and some model selection methods
use these with asymptotic expansions and approximations for their
derivation or justification. There are various ways to think
about
asymptotics in this problem. The simplest\vadjust{\goodbreak} is to let $n \rightarrow
\infty$ in such a way that various matrices (such as $n^{-1}\XX^T\VV^{-1}\XX$)
converge to positive definite limits. For independent
cluster models, the standard methods are to allow the number of
independent groups or clusters $m \rightarrow\infty$ with either $\max
(n_i)$ bounded or $\min(n_i) \rightarrow\infty$. In this model, the
case of $m$ fixed and $\min(n_i) \rightarrow\infty$ is only useful if
$\PS$ is known because otherwise $\PS$ cannot be estimated
consistently. 
Most methods also impose further restrictions on the dimensions of the
model. The usual fixed parameter case has $p+q\ll n$, although some
estimation methods even require $p+q+s \ll n$.

\section{Information Criteria} \label{secinformation}

Information criteria are widely used to compare and select models. In
practice, they are applied by finding the model that minimizes an
estimate of a criterion that is generally of the form $Q_M(\hat{\bth
}_M) + \alpha_n(d_M)$, where $Q_M$ is a loss function which, for
candidate models $M_1$ and $M_2$ satisfying $M_1 \subset M_2$,
satisfies $Q_{M_2}(\hat{\bth}_{M_2}) \le Q_{M_1}(\hat{\bth}_{M_1})$ (it
is often minus twice the log-likelihood or a closely related function)
and the penalty function $\alpha_n$ is a function of the model
complexity $d_M$. There are a number of approaches to obtaining
information criteria such as the Akaike approach, Schwarz's Bayesian
approach, etc. and within these there can be multiple possible
criteria. For example, for the linear mixed model (\ref{eqremodel}) to
define the loss function we can use the log-likelihood, the conditional
log-likelihood or the\break REML criterion and for the transformation model
(\ref{eqtrmodel}) we can use the log-likelihood or the REML criterion.
For the linear regression model, $\alpha_n$ is often just a function of
the number of parameters in the model (which in the present context is
$p+q$; see M\"uller and Welsh, \citeyear{Mueller2010}, for a review) but for linear mixed
models can be more complicated.

The Akaike Information (\cite{Akaike1973}) is a measure of the ability of a
model fitted using a particular estimator to predict an independent
copy of the observed data. The particular measure used is the
expectation over both the data and the independent copy of the data, of
minus twice the logarithm of a density-like function representing the
model which is evaluated at the independent copy of the data and the
estimator of the unknown parameters based on the data. This definition
is of necessity vague because we can define different versions of the
Akaike Information using different log density-like functions and we
can consider various estimators of $\bth$ in these functions. In
particular, if we let $\hat{\bth}(\yy)$ be an estimator of $\bth$ based
on the data $\yy$, $\XX$ and $\ZZ$, and let $\yy^*$ be an independent
copy of $\yy$, then the marginal Akaike Information for a class of
distributions with density-like function $g(\yy; \bth)$ is $-2\E_{\yy}\E_{\yy^*} \log[g\{\yy^*; \hat{\bth}(\yy)\}]$ and the conditional Akaike
Information for a class of distributions with conditional\vspace*{1pt} (i.e., $\yy
|\uu$) density-like function $f(\yy; \bth, \uu)$ is $-2\E_{\yy,\uu}\E_{\yy^*|\uu} \log[f\{\yy^*; \hat{\bth}(\yy), \hat{\uu}(\yy)\}]$, where
$\hat{\uu}(\yy)$ is a predictor of $\uu$. The expectations in the
marginal case are taken with respect to either the linear mixed model
(\ref{eqremodel}) or the transformation model (\ref{eqtrmodel}) and
in the conditional case they are taken with respect to the linear mixed
model (\ref{eqremodel}). The marginal Akaike Information (based on the
log-likelihood or the REML criterion) is meaningful when the
independent copy of the data $\yy^*$ is independent of $\yy$; the
conditional criterion (based on the conditional log-likelihood) is
meaningful for the linear mixed model (\ref{eqremodel}) when $\yy^*$
and $\yy$ are conditionally independent given $\uu$ so the same random
effects are common to $\yy$ and $\yy^*$.

The model chosen from the specified class is a model that minimizes an
estimator called the Akaike Information Criterion (AIC) of the Akaike
Information. Depending on how we define the Akaike Information, it is
natural to consider estimating it using minus twice the corresponding
log-likelihood, REML criterion or conditional log-likelihood. These
functions are biased estimators of the Akaike Information because they
use the same observed data $\yy$ both to compute the parameter
estimator and to evaluate the function itself. We can evaluate the bias
and try to make an approximate adjustment for it: The penalty $\alpha_n(d_M)$
in an AIC can be interpreted as an adjustment to reduce bias.
Much of the focus in the literature is on adjusting the bias to obtain
a good estimator of the Akaike Information, although this is not the
real problem in model selection.\looseness=1

Model selection methods like AIC which make use of the log-likelihood
are closely related to likelihood ratio tests in which the models
correspond to different hypotheses, a relationship which implies that
hypothesis tests can be used to suggest new criteria. The important
Bayesian Information Criterion or BIC (\cite{Schwarz1978}) can be derived as
an approximation to the Bayes factor for testing two hypotheses $M_0$
and $M_1$ or from asymptotic arguments to construct criteria which lead
to consistent model selection.

There are a number of other information criteria in the literature.
They are derived for various reasons from various considerations. Some
of them are simply general criteria which could be applied in the
linear mixed model, others have been applied to or developed for the
linear mixed model. It is noteworthy that these are mainly marginal
criteria (i.e., based on the log-likelihood) and that there are not
many proposals outside the AIC framework for conditional criteria
(i.e., based on the conditional log-likelihood). We discuss AIC, BIC
and some of the other criteria in the subsections that follow.

\subsection{Marginal AIC}\label{marginalAIC}

The most widely used AIC criterion is what \citet{Vaida2005}
call the marginal AIC criterion, namely,
%
\begin{equation}
\label{eqmAIC} m\mathit{AIC} = -2 \ell(\hat{\bth}) + 2a_n(p + q),
\end{equation}
where $a_n=1$ or $a_n = n/(n-p-q-1)$ in the finite sample form
(\cite{Sugiura1978}). For the independent cluster model, $m\mathit{AIC}$ is
asymptotically equivalent to leave-one-cluster-out cross-validation
using a marginal generalized least squares criterion (\cite{Fang2011}); see
Section~\ref{secother}. The R function \texttt{lme()} uses $m\mathit{AIC}$ with
$a_n = 1$ and SAS \texttt{Proc Mixed} uses both the asymptotic and the
finite sample forms.

The marginal AIC represents the application of a general theory to the
linear mixed model (\ref{eqremodel}) or the transformation model (\ref
{eqtrmodel}) without taking into account the specific nature of these
models. For the linear mixed model (\ref{eqremodel}) with $\PS= \gamma^2 \bA^T\bA$, $\bA$
known and $\SI= \sigma^2\II_n$ so $q=2$, \citet{Greven2010} show that $m\mathit{AIC}$ is positively biased for the marginal
Akaike Information, where the bias depends on the unknown variance
parameter $\gamma^2$ and does not vanish asymptotically if $\gamma^2=0$.
This means that there is no simple bias correction to make
$m\mathit{AIC}$ exactly unbiased and the fact that the variance parameters are
constrained by the boundary of the parameter space ought to be built
into the penalty. A further issue with $m\mathit{AIC}$ is that the model
complexity term in the penalty $p+q$ gives the same weight to the
parameters in $\be$ and the parameters in $\ta$. There is no obvious
reason why this should be the case; in the variance component model,
each parameter $\gamma_k$ in $\ta$ represents the variance of $r_k$
random effects so is absorbing $r_k$ other estimates and arguably
should be given a greater weight to reflect this. The precise form of a
penalty giving different weight to the parameters in $\be$ and $\ta$ is
not clear. However, it is possible that using a different criterion to
minus twice the log-likelihood may have the effect of rescaling the
parameters so that it then makes sense to give them equal weight in the penalty.

\citet{Shang2008} propose using the bootstrap to estimate the
appropriate bias-adjust\-ment or penalty for marginal AIC. Let $\{\yy^*_b\dvtx b=\break1,\ldots,B\}$ be a set of $B$ bootstrap samples of $\yy$, let
$\{\ell_b^*(\bth)\dvtx b=1,\ldots,B\}$ denote the log-likelihoods for each
of the $B$ bootstrap samples [i.e., $\ell_b^*(\bth)$ is the
log-likelihood for $\bth$ computed using $\yy^*_b$] and let $\{\hat{\bth
}{}^*_b\dvtx b=1,\ldots,B\}$ denote the maximum likelihood estimates for each
bootstrap sample. Then Shang and Cavanaugh propose the bootstrap AIC criteria
\[
m\mathit{AIC}_{B1} = -2 \ell(\hat{\bth}) - \frac{2}{B}\sum
_{b=1}^B \bigl\{\ell\bigl(\hat {
\bth}^*_b\bigr) - \ell_b^*\bigl(\hat{
\bth}^*_b\bigr)\bigr\}
\]
and
\begin{eqnarray*}
m\mathit{AIC}_{B2} &=& -2 \ell(\hat{\bth}) - \frac{4}{B}\sum
_{b=1}^B \bigl\{\ell\bigl(\hat {
\bth}^*_b\bigr) - \ell(\hat{\bth})\bigr\}\\
& =& 2 \ell(\hat{\bth}) -
\frac{4}{B}\sum_{b=1}^B \ell\bigl(
\hat{\bth}{}^*_b\bigr).
\end{eqnarray*}
In their simulations, Shang and Cavanaugh use the parametric bootstrap
but other types of bootstrap could be used.

Rather than applying general results to the specific context,
Srivastava and Kubokawa (\citeyear{Srivastava2010}) obtain a different criterion by working
directly within the linear mixed model (\ref{eqremodel}) with $\SI
=\sigma^2\II_n$. Treating $\PS/\sigma^2$ as known (so there are $p+1$
unknown parameters $\be$ and $\sigma^2$), they obtain the criterion
%
\begin{equation}
\label{eqmAICsk} m\mathit{AIC}_{\mathit{SK}} = -2 \ell(\hat{\bth}) +
\frac{2n(p + 1)}{n-p-2}.
\end{equation}
(They do not assume that $\XX$ is of full rank so their expression has
$\rank{(\XX)}$ in place of $p$.) When $\PS/\sigma^2$ is unknown, they
replace it by an estimator without any further adjustment for
estimating these additional parameters. There seems little reason to
expect the criterion to perform well in this case, unless the number of
additional variance parameters $q_{\gamma}$ is small.

The REML criterion function $\ell_R(\ta)$ is a modified profile
likelihood for $\ta$ so is not a function of $\be$. This seems to imply
that $\ell_R(\ta)$ may not be useful for selecting regression
parameters. It is, however, an implicit function of $\be$ in the sense
that we need to specify an $\XX$ when we do the profiling and different
choices of $\XX$ correspond to different choices of $\be$ which change
the value of $\ell_R(\ta)$. This means that we can in fact consider
using a version of marginal AIC based on the REML criterion function
for model selection. SAS \texttt{Proc Mixed} uses
%
\begin{equation}
\label{eqmAICR} m\mathit{AIC}_R = -2 \ell_R(\hat{\ta}) +
2a_n^*q,
\end{equation}
with $a_n^*=(n-p)/(n-p-q-1)$. The form of $a_n^*$ is related to the
second order adjustment $a_n$ of \citet{Sugiura1978} after adjusting the
sample size for implicitly having estimated $\be$. There is no other
penalty for implicitly having estimated $\be$; this is the antithesis
of the Srivastava--Kubokawa criterion (\ref{eqmAICsk}) which does not
adjust the complexity in the penalty $p+1$ for estimating $\PS/\sigma^2$,
because it does not adjust the complexity $q$ for estimating $\be$
and it is unclear how well this criterion performs.

\citet{Kubokawa2011} considers using marginal AIC with minus twice the
log-likelihood evaluated at the generalized least squares estimator $\be
(\ta) = \break (\XX^T\VV^{-1}\XX)^{-1}\XX^T\VV^{-1}\yy$ of $\be$ and a general
consistent estimator $\hat{\ta}$ of $\ta$ which admits an expansion of
the form
%
\begin{equation}
\label{eqtauexpansion} \hat{\ta} - \ta= \bt_1(\ta) +
\bt_2(\ta) + O_p\bigl(n^{-3/2}\bigr),
\end{equation}
where  $\E\{\bt_1(\ta)\} = \E\{(t_{11}(\ta),\ldots, t_{1q}(\ta))^T\} =
\zer$,\break $\bt_1(\ta) = O_p(n^{-1/2})$ and $\bt_2(\ta) = (t_{21}(\ta
),\ldots,\break  t_{2q}(\ta))^T = O_p(n^{-1})$. He suggests replacing the
penal\-ty in (\ref{eqmAIC}) by $2\{p+h_m(\hat{\ta})\}$, where
\[
h_m(\ta) =\frac{1}{2}\sum_{k=1}^q
\E \biggl[\trace \biggl\{\frac{\partial\VV
}{\partial\tau_i} \frac{\partial^2 t_{1i}(\ta)}{\partial\yy\,\partial\yy^T} \biggr\} \biggr].
\]
This shows the effect of using different estimators and confirms that
the AIC approach depends both on the model and the estimator used to
fit it. When maximum likelihood or REML are used to estimate $\ta$, the
penalty reduces back to $2(p+q)$ and we obtain $m\mathit{AIC}$ defined in (\ref
{eqmAIC}).

\subsection{Conditional AIC}\label{condAIC}

The conditional Akaike Information is defined only for the linear mixed
model (\ref{eqremodel}) and not for the transformation model (\ref
{eqtrmodel}). We need to predict $\uu$ or, equivalently and more
conveniently, $\vv=\GA\uu$ as well as estimate the parameters $\bth$ so
there are $p+q+s$ unknown quantities to estimate. When the variance
parameters $\ta$ are known, $\vv$ is often predicted using the best
linear unbiased predictor (BLUP)
%
\begin{equation}
\label{eqblup} \hat{\vv}(\ta) = \PS\ZZ^T\VV^{-1}\bigl\{
\yy- \XX\hat{\be}(\ta)\bigr\},
\end{equation}
where $\hat{\be}(\ta)$ is the generalized least squares estimator
defined in (\ref{eqgls}). When $\ta$ is unknown, we use an estimated
BLUP or EBLUP $\hat{\vv} = \hat{\vv}(\hat{\ta})$. Since we are working
with $\vv$ rather than $\uu$ in this section, it is convenient to treat
the conditional log-likelihood (\ref{condlogliktheta}) as a function
of $\vv$ rather than $\uu$.

The generalized least squares estimator $\hat{\be}(\ta)$ and the BLUP
$\hat{\vv}(\ta)$ of $\vv$ can be obtained as the solution of
Henderson's (\citeyear{Henderson1950}) mixed model equations
\begin{eqnarray*}
&&\pmatrix{ \XX^T\SI^{-1}
\XX& \XX^T\SI^{-1}\ZZ
\vspace*{2pt}\cr
\ZZ^T\SI^{-1}\XX& \ZZ^T\SI^{-1}\ZZ+
\PS^{-1} }
\pmatrix{\hat{\be}(\ta)
\vspace*{2pt}\cr
\hat{\vv}(\ta) } \\
&&\quad= \pmatrix{ \XX^T
\vspace*{2pt}\cr
\ZZ^T }\SI^{-1}\yy.
\end{eqnarray*}
These equations enable us to write $\XX\hat{\be}(\ta) + \ZZ\hat{\vv}(\ta
) = \HH_1(\ta) \yy$, where
\begin{eqnarray*}
\HH_1(\ta) &=& (\XX, \ZZ) \pmatrix{
\XX^T\SI^{-1}\XX& \XX^T\SI^{-1}\ZZ
\vspace*{2pt}\cr
\ZZ^T\SI^{-1}\XX& \ZZ^T\SI^{-1}\ZZ+
\PS^{-1} }^{-1}\\
&&{}\cdot \pmatrix{ \XX^T\SI^{-1}
\vspace*{2pt}\cr
\ZZ^T\SI^{-1} },
\end{eqnarray*}
and then to treat $\HH_1(\ta)$ as a ``hat'' matrix. In particular, when
$\ta$ is known,
%
\begin{eqnarray}
\label{eqedf} \rho(\ta) &=& \trace\bigl\{\HH_1(\ta)\bigr\} \nonumber\\
&=& \trace
\bigl[\bigl(\XX^T\VV^{-1}\XX\bigr)^{-1}
\XX^T\VV^{-1}\SI\VV^{-1}\XX\bigr] + n\\
&&{} - \trace
\bigl(\SI\VV^{-1}\bigr)\nonumber
\end{eqnarray}
is the \textit{effective degrees of freedom} used in estimating $\be$
and $\vv$ (\cite{Hodges2001}). The effective degrees of freedom
satisfies $p \le\rho(\ta) \le p+s$ so lies between the degrees of
freedom of the regression model without $\vv$ and the regression model
treating $\vv$ as fixed effects (\cite{Vaida2005}). Computing
$\HH_1(\ta)$ requires both $\SI^{-1}$ and $\PS^{-1}$ but (\ref{eqedf})
shows that computing $\rho(\ta)$ only requires $\VV^{-1}$, which should
be more stable.

We have defined the effective degrees of freedom for the general model
(\ref{eqremodel}). It is worth noting that most of the literature on
conditional AIC actually assumes that $\SI=\sigma^2\II_n$ so $\de$
consists of just the scalar parameter $\sigma$. In this case, it is
convenient to let $\PS_{*}=\PS/\sigma^2$ and $\VV_{*}= \ZZ\PS_{*}\ZZ^T
+ \II_n$. It follows that $\VV= \sigma^2\VV_{*}$ and hence that $\hat
{\be}(\ta)$, $\hat{\vv}(\ta)$ and $\rho(\ta)$ are functions of the
parameters in $\PS_{*}$. Some conditional AIC are derived by treating
$\PS_{*}$ as known and subsequently replacing it by an estimator. This
case is subsumed within the general notation so we handle it by drawing
specific attention to it when necessary rather than by introducing
additional notation.

\begin{table*}
\caption{Penalties $\alpha_n(\hat{\bth})$ for conditional AIC. The
entries in the table are $\alpha_n(\hat{\bth})/2$ so the actual penalty
is obtained\break by multiplying each entry by $2$. All the criteria other
than the two asterisked criteria (Burnham and White, Kubokawa) assume
$\SI=\sigma^2\II_n$. The references are given in an abbreviated form
using the first letters of the authors' names\break and the last two digits
of the publication date}\label{tblcAICpenalties}
\begin{tabular*}{\textwidth}{@{\extracolsep{\fill}}lcc@{}}
\hline
\textbf{Notation} & $\bolds{\alpha_n(\hat{\bth})/2}$ & \textbf{Reference}\\
\hline
Maximum likelihood (\ref{eqcAIC}) \\
\quad$c\mathit{AIC}_{\mathit{BW}}$ & $\rho(\hat{\ta})+q$ & BW02$^*$ \\
\quad$c\mathit{AIC}_{\mathit{VB}}$ & $\rho(\hat{\ta})+1$ & VB05\\
& $\frac{n}{n-p-2}\{\rho(\hat{\ta})+1 - \frac{\rho(\hat{\ta})-p}{n-p}\}
$ & VB05\\
\quad$c\mathit{AIC}_{\mathit{LWZ}}$ & $\trace(\partial\hat{\yy}/\partial\yy) $& LWZ08 \\
& $+\hat{\sigma}^2 (\hat{\yy}-\yy)^T \frac{\partial\hat{\sigma
}^{-2}}{\partial\yy} + \frac{1}{2}\hat{\sigma}^4 \trace ( \frac
{\partial^2 \hat{\sigma}^{-2}}{\partial\yy\,\partial\yy^T} )$ &
(Tech. rep.)\\
\quad$c\mathit{AIC}_{{\mathit{GK}}}$ & $\hat{\nu}(\hat{\sigma}^2) + 1$ & GK10\\[6pt]
REML (\ref{eqcAICR}) \\
\quad$c\mathit{AIC}_{R,\mathit{VB}}$ & $\rho(\hat{\ta}_R)+1$ & VB05\\
& $\frac{n-p-1}{n-p-2}\{\rho(\hat{\ta}_R)+1 - \frac{p+1}{n-p-1}\}$ &
VB05 \\
\quad$c\mathit{AIC}_{R,{\mathit{GK}}}$ & $\hat{\nu}_R(\hat{\sigma}_R^2) + 1$ & GK10\\[6pt]
Specific $\tilde{\sigma}_0^2$ (\ref{eqSKcAIC})\\
\quad$c\mathit{AIC}_{\mathit{SK}}$ & $\frac{n[\trace\{(\XX,\ZZ)\CC(\hat{\ta})\} + 1]}{n-\rank\{
(\XX, \ZZ)\}-2}$ & SK10 \\
General $\hat{\ta}$\\
\quad$c\mathit{AIC}_{K}$ & $\rho(\hat{\ta}) + \hat{h}_c(\hat{\ta})$ & K11$^*$ \\
\hline
\end{tabular*}
\end{table*}

Conditional AIC criteria are constructed using minus twice the
conditional log-likelihood as the loss\vadjust{\goodbreak} function plus a penalty. All the
proposed criteria are of the form
%
\begin{equation}
\label{eqcAIC} c\mathit{AIC}_{\alpha_n} = -2 \ell(\hat{\bth}| \hat{\vv}) +
\alpha_n(\hat{\bth})
\end{equation}
with different estimators $\hat{\bth}$, predictors $\hat{\vv}$ and
different penalties $\alpha_n(\hat{\bth})$. A summary of the proposed
penalties $\alpha_n(\hat{\bth})$ is given in Table~\ref{tblcAICpenalties}; we discuss each of these choices in turn.

\citet{Burnhamb2002} and \citet{Burnhama2002} propose using
the function
\[
\label{eqcAICba} \alpha_{n,\mathit{BW}}(\hat{\bth}) = 2\bigl\{\rho(\hat{\ta})
+ q\bigr\}.
\]
Here $\rho(\hat{\ta})$ measures the effect of estimating $\be$ and~$\vv
$; $q$ is included to try to accommodate the effect of estimating $\ta
$. For the case that $\SI=\sigma^2\II_n$, \citet{Vaida2005}
instead suggest using
%
\begin{eqnarray}
\label{eqcAICvb} &&\alpha_{n,\mathit{VB}}(\hat{\bth})
\nonumber
\\[-8pt]
\\[-8pt]
\nonumber
&&\quad= \frac{2n}{n-p-2}\biggl
\{\rho(\hat{\ta}) + 1 - \frac{\rho(\hat{\ta}) - p}{n-p} \biggr\}.
\end{eqnarray}
The function $\alpha_{n,\mathit{VB}}(\hat{\bth})$ is asymptotic to \mbox{$2\{\rho(\hat
{\ta}) + 1\}$}, as $n \rightarrow\infty$ with $p$, $q$ fixed, which is
the effective degrees of freedom for estimating $\be$ and $\vv$ plus
one degree of freedom for estimating $\sigma^2$. Vaida and Blanchard
derive this penalty assuming that $\PS_{*}$ is known and point out
that, in this case, their criterion is the same as the DIC of
Spiegelhalter et al. (\citeyear{Spiegelhalter2002}); see Section~\ref{secBayes}. For the
independent cluster model, $c\mathit{AIC}$ is asymptotically equivalent to
leave-one-observation-out cross-validation with a conditional least
squares criterion (\cite{Fang2011}); see Section~\ref{secother}. When $\PS_{*}$
is unknown, Vaida and Blanchard suggest simply using the
estimated version (\ref{eqcAICvb}), that is, without adjustment for
estimating $\PS_{*}$. Their argument is that $\ell(\bth| \vv)$ does not
depend on $\PS_{*}$. This is plausible with $\vv= \GA\uu$ absorbing $\GA
$ into $\uu$ but $\hat{\be}(\ta)$ and $\hat{\vv}(\ta)$ are functions of
$\PS_{*}$.

In a technical report accompanying their paper, Liang, Wu and Zou (\citeyear{Liang2008})
propose a different penal\-ty to take the estimation of $\PS_{*}$ into
account. When $\SI=\sigma^2\II_n$, they propose using
\begin{eqnarray*}
\label{eqcAIClwz} \alpha_{n,\mathit{LWZ}}(\hat{\bth}) &=& 2 \biggl\{\trace \biggl(
\frac{\partial\hat{\yy
}}{\partial\yy} \biggr) + \hat{\sigma}{}^2 (\hat{\yy}-
\yy)^T \frac{\partial
\hat{\sigma}{}^{-2}}{\partial\yy} \\
&&\hspace*{47pt}{}+ \frac{1}{2}\hat{
\sigma}^4 \trace \biggl( \frac{\partial^2 \hat{\sigma}{}^{-2}}{\partial\yy\,\partial\yy^T} \biggr) \biggr\},
\end{eqnarray*}
where $\hat{\yy}=\XX\hat{\be} + \ZZ\hat{\vv}$. When $\sigma^2$ is
known, the penalty reduces to just the first term $\trace(\partial\hat
{\yy}/\partial\yy)$, which is the \textit{generalized degrees of
freedom} of \citet{Ye1998}. \citet{Greven2010} derive analytic
representations for these penalties. Let $\hat{\nu}(\sigma^2)$ denote
the analytic representation of the generalized degrees of freedom when
$\sigma^2$ is known. On the basis of simulations, Greven and Kneib
suggest using $\hat{\nu}(\hat{\sigma}^2)+1$ when $\sigma^2$ is unknown.
Their penalty in the general case is therefore
\[
\label{eqcorrectedcAIC} \alpha_{n,{\mathit{GK}}}(\hat{\bth}) = 2\bigl\{\hat{\nu}
\bigl(\hat{\sigma}{}^2\bigr)+1\bigr\}.
\]
The expression for $\hat{\nu}(\sigma^2)$ is quite complicated because
it explicitly allows for the variance parameters to lie on the boundary
of the parameter space. The penalty has been implemented in R and code
is available from the online Supplementary Material for the paper.

\citet{Vaida2005} and \citet{Greven2010} also consider
fitting the linear mixed model with $\SI=\sigma^2\II_n$ using the REML
estimator $\hat{\ta}_{R}$ and then $\hat{\be}_R = \hat{\be}(\hat{\ta
}_{R})$ and $\hat{\vv}_R = \hat{\vv}(\hat{\ta}_{R})$ with $\hat{\be}(\ta
)$ and $\hat{\vv}(\ta)$ defined in (\ref{eqgls}) and (\ref{eqblup}),
respectively. Note that they use the same conditional likelihood as in
the definition of $c\mathit{AIC}$ (\ref{eqcAIC}) evaluated at the $\hat{\be}_R$
and $\hat{\vv}_R$ but with a different penalty. Thus, the criteria are
of the form
%
\begin{equation}
\label{eqcAICR} c\mathit{AIC}_{R,\mathit{VB}}(\hat{\bth}_R) =-2 \ell(\hat{
\bth}_R| \hat{\vv}_R) + \alpha_n(\hat{
\bth}_R).
\end{equation}
\citet{Vaida2005} propose the penalty
\begin{eqnarray*}
&&\alpha_{n,R,\mathit{VB}}(\hat{\bth}_R)\\
&&\quad = \frac{2(n-p-1)}{n-p-2} \biggl\{
\rho(\hat {\ta}_{R}) + 1 + \frac{p+1}{n-p-1} \biggr\}.
\end{eqnarray*}
This penalty is asymptotic to $2\{\rho(\hat{\ta}_{R}) + 1\}$, as $n
\rightarrow\infty$ with $p$, $q$ fixed, which is like their penalty
for the maximum likelihood estimator. \citet{Greven2010} also
derive an analytic representation $\hat{\nu}_R(\sigma^2)$ for the Liang, Wu and Zou
(\citeyear{Liang2008}) penalty for\break the REML estimator. It turns out that $\hat
{\nu}_R(\sigma^2)$ is \mbox{different} from $\hat{\nu}(\sigma^2)$ for the
maximum likelihood\break \mbox{estimator}. The penalty in this case is
therefore\break
$\alpha_{n,R,{\mathit{GK}}}(\hat{\bth}_R)=2\{\hat{\nu}_R(\sigma^2)+1\}$.

Srivastava and Kubokawa (\citeyear{Srivastava2010}) derive other conditional criteria by
changing the estimators of the parameters at which minus twice the
conditional log-likelihood is evaluated and then adjusting the penalty
appropriately. For the model with $\SI=\sigma^2\II_n$ and $\PS_{*}$
known, Srivastava and Kubokawa propose replacing the maximum likelihood\vspace*{1pt}
estimator $\hat{\sigma}{}^2 = (\yy-\XX\hat{\be})^T\hat{\VV}{}^{-1}_{*}(\yy
-\XX\hat{\be})/n$ of $\sigma^2$ by the estimator
\[
\tilde{\sigma}_0^2 = \bigl\{\yy- (\XX,\ZZ)\tilde{\xi}
\bigr\}^T\bigl\{\yy-( \XX, \ZZ )\tilde{\xi}\bigr\}/n,
\]
where $\tilde{\xi} = \{( \XX, \ZZ)^T( \XX, \ZZ)\}^{+}( \XX, \ZZ)^T\yy$
and $\bA^{+}$ is the Moore--Penrose inverse of $\bA$. This change in the
variance estimator involves treating $\uu$ as an unknown, fixed
parameter which is to be estimated, here by ordinary least squares. The
idea of treating $\uu$ in this way is used by \citet{Jiang2003}
(Section~\ref{secother}). The use of $\tilde{\sigma}_0^2$ changes the
form of the penalty. For any estimators $\hat{\be}_C$ and $\hat{\vv}_C$
satisfying $(\hat{\be}{}^T_C, \hat{\vv}{}^T_C)^T =\CC(\ta)\yy$, they obtain
the modified conditional criterion
%
\begin{eqnarray}
\label{eqSKcAIC} c\mathit{AIC}_{\mathit{SK}} &=& -2 \ell\bigl(\hat{\be}_C,
\tilde{\sigma}_0^2| \hat{\vv}_C\bigr)
\nonumber
\\[-8pt]
\\[-8pt]
\nonumber
&&{}+
\frac{2n[\trace\{(\XX,\ZZ)\CC(\hat{\ta})\} + 1]}{n-\rank\{(\XX, \ZZ)\}-2}.
\end{eqnarray}
Note that here the parameters $\ga$ are absorbed into $\hat{\vv}_C$ so
do not appear separately in the conditional log-likelihood. When $\hat
{\be}_C$ is either the maximum likelihood or the least squares
estimator and $\hat{\vv}_C$ is the BLUP $\hat{\vv}$, $\trace\{(\XX,\ZZ
)\CC(\ta)\}=\rho(\ta)$; when $\hat{\be}_C$ and $\hat{\vv}_C$ are the
least squares estimators extracted from~$\tilde{\xi}$, $\trace\{(\XX,\ZZ
)\CC(\ta)\}=\rank(\XX,\ZZ)$. In the first case, the penalty is the
asymptotic version of the Vaida--Blanchard penalty (\ref{eqcAICvb})
with $p$ replaced by the larger number $\rank\{(\XX, \ZZ)\}$ so the
Srivastava--Kubokawa penalty is larger than the asymptotic
Vaida--Blan\-chard penalty. When $\PS_{*}$ is unknown, for other
estimators which use $\PS_{*}$, Srivastava and Kubokawa (\citeyear{Srivastava2010}) propose
replacing it by an estimator $\hat{\PS}_{*}$. For computational
reasons, they consider using the truncated method of moments estimators
for the special cases, but any consistent estimator can be used.

For the linear mixed model with a general $\SI$, Kubokawa (\citeyear{Kubokawa2011})
considers estimators $\hat{\ta}$ of $\ta$ which satisfy the second
order expansion (\ref{eqtauexpansion}). Let\break $\dd\{f(\ta)\} =
(\partial f(\ta)/\partial y_j)$ denote the $n$-vector of deri\-vatives of
$f$ with respect to $\yy$ and $\DD\{f(\ta)\} =\break (\partial^2 f(\ta
)/\partial y_j\, \partial y_k)$ denote the $n \times n$ matrix of
second~derivatives of $f$ with respect to $\yy$. Then,\break under the condition
that~the three terms\break $\E (\trace[\DD\{t_{2i}(\ta)\}] )$, $\E
(\trace[\DD\{t_{1i}(\ta)\}t_{1j}(\ta)] )$ and $\E (\trace[\dd\{
t_{1i}(\ta)\}\dd\{t_{1j}(\ta)\}^T] )$ are all $O(n^{-1})$, Ku\-bokawa (\citeyear{Kubokawa2011}) derives the penalty
\[
\label{eqKcAIC} \alpha_{n,K}(\hat{\bth}) = 2\bigl\{\rho(\hat{\ta}) +
\hat{h}_c(\hat{\ta})\bigr\},
\]
where $\hat{h}_c(\ta)$ is an estimator of
\begin{eqnarray*}
h_c(\ta) & = & -\frac{1}{2}\sum_{i=1}^q
\trace \biggl\{\biggl(\frac{\partial\SI}{\partial\tau_i} - 2\SI\VV^{-1}\frac{\partial\VV
}{\partial\tau_i}
\biggr) \\
&&\hspace*{89pt}{}\cdot\E\bigl[\DD\bigl\{t_{1i}(\ta)\bigr\}\bigr] \biggr\} \\
&&{} - \sum
_{i=1}^q \trace \biggl\{\frac{\partial\SI}{\partial\tau_i}\bigl(
\SI^{-1} - \VV^{-1}\bigr) \biggr\}\E\bigl\{t_{2i}(
\ta)\bigr\}
\\
&&{} - \sum_{i=1}^q \sum
_{j=1}^q \trace \biggl\{\frac{1}{2}
\frac{\partial^2
\SI}{\partial\tau_i\,\partial\tau_j}\bigl(\SI^{-1} - \VV^{-1}\bigr)\\
&&\hspace*{62pt}\quad{} +
\frac
{\partial\SI}{\partial\tau_i}\biggl(\frac{\partial\SI^{-1}}{\partial\tau_j} - \frac{\partial\VV^{-1}}{\partial\tau_j}\biggr) \biggr\}\\
&&\hspace*{44pt}{}\cdot\E
\bigl\{t_{1i}(\ta )t_{1j}(\ta)\bigr\},
\end{eqnarray*}
which is obtained by replacing all the unknown quantities by estimators.
It is a considerable task to derive the second order expansion (\ref
{eqtauexpansion}) and then to derive the expressions which are needed
to compute $h_c(\ta)$, but \citet{Kubokawa2011} provides results for the
maximum likelihood and REML estimators. These are still quite
complicated for general use so \citet{Kubokawa2011} specializes the
expressions further to three particular models, namely, the variance
component model (\ref{eqremodelvc}), the random intercept regression
model and the Fay--Herriot model.

\subsection{BIC and Schwarz Criteria} \label{secbic}

The simplest and most widely used BIC for the linear mixed model (\ref
{eqremodel}) or the transformation model (\ref{eqtrmodel}) is
obtained by taking the marginal AIC (\ref{eqmAIC}) and replacing the
constant $2$ in the penalty by $\log(n)$ to obtain
\[
\BIC = -2\ell(\hat{\bth}) + \log(n) (p+q).
\]
This is the definition used by \texttt{lme()} in R and by SAS \texttt
{Proc Mixed}. This definition ensures that $\BIC$ bears the same
relationship to $m\mathit{AIC}$ for model (\ref{eqremodel}) as BIC bears to AIC
in regression and so should inherit some of its properties.
Specifically, the increased weight in the penalty should encourage
$\BIC$ to select smaller models than $m\mathit{AIC}$. Obviously,\break other $m\mathit{AIC}$ can
be converted to $\BIC$ in the same way by multiplying the $m\mathit{AIC}$ penalty
by $\log(n)/2$.


A more sophisticated approach is possible if we re-examine the
relationship between BIC and the Bayes factor. After reordering if
necessary, partition $\bth= (\bth_0^T, \bth_1^T)^T$ into $\bth_0 \in
R^{r_0}$, $\bth_1 \in R^{r_1}$, $p + q = r_0+r_1$, and consider
comparing the model $M_0\dvtx \bth_0 = \bth_{00}$ with $M_1\dvtx \bth_0 \ne\bth_{00}$.
Let $h_0$ be the prior density for $\bth_1$ under $M_0$ and let
$h_1$ be the prior density for $\bth$ under $M_1$. Then the Bayes
factor for comparing $M_0$ to $M_1$ is the ratio of the posterior odds
to the prior odds for a model
%
\begin{eqnarray}
\label{eqbayesfac}&& \frac{\Pr(M_0|\yy)/\Pr(M_1|\yy)}{\Pr(M_0)/\Pr(M_1)}
\nonumber
\\[-8pt]
\\[-8pt]
\nonumber
&&\quad= \frac{\int g(\yy
|\bth_{00},\bth_1)h_0(\bth_1) \,d\bth_1}{\int g(\yy|\bth_0, \bth_1)h_1(\bth_0, \bth_1) \,d\bth_0\,d\bth_1},
\end{eqnarray}
where $g(\yy|\bth) = \exp\{\ell(\bth)\}$ is the marginal likelihood of
the model. If we hold $M_0$ constant at say the simplest model under
consideration, this leads to choosing the model that minimizes $-2\log\{
\int g(\yy|\bth_0, \bth_1)\cdot  h_1(\bth_0, \bth_1) \,d\bth_0\,d\bth_1\}$. BIC
can be obtained using Lap\-lace's method to approximate the integral in
this expression.

\citet{Pauler1998} uses this approach to derive a\break Schwarz criterion to
select the regression parameter $\be$ in the independent cluster model.
Partition $\be= (\be_0^T, \be_1^T)^T$ into $\be_0 \in R^{p_0}$, $\be_1
\in R^{p_1}$, $p=p_0+p_1$, and consider testing the null hypothesis
$M_0\dvtx \be_0 = \be_{00}$ against $M_1\dvtx \be_0 \ne\be_{00}$. Pauler
required $\be_0$ to be null orthogonal to $(\be_1^T,\ta^T)^T$ and, if
the prior density for $M_0$ is $h_0(\be_1, \ta)$, the prior for $M_1$
to be of the form $h_1(\be, \ta) = h_0(\be_1, \ta)h(\be_0|\be_1,\ta)$.
She notes that if $\uu$ and $\ee$ are Gaussianly distributed, $\be$ and
$\ta$ are orthogonal (the information matrix is block diagonal) and
that $\be_1$ can be made null orthogonal to $\be_0$ by transforming $
\be_1 \rightarrow\be_1 + (\XX_1^T\VV^{-1}\XX_1)^{-1}\XX_1\VV^{-1}\XX_0\be_0$,
where $\XX= (\XX_0, \XX_1)$ is partitioned conformably\break with~$\be$.
Then, using Laplace's method, she approximates the Bayes factor
for comparing $M_0$ to $M_1$ by
%
\begin{eqnarray}\qquad
\label{eqschwarz} S &=& \ell\bigl\{\hat{\bth}(\be_{00})\bigr\} - \ell(
\hat{\bth}) -\tfrac{1}{2}p_0 \log (2\pi)
\nonumber
\\[-8pt]
\\[-8pt]
\nonumber
&&{} + \tfrac{1}{2}
\log\bigl|\XX^T_0 \hat{\VV}{}^{-1}\XX_0\bigr|
- \log\bigl\{h(\hat {\be}_0|\hat{\be}_1, \hat{\ta})
\bigr\},
\end{eqnarray}
where $\hat{\bth}$ is the maximum likelihood estimator of $\bth$ and
$\hat{\bth}(\be_{00})$ maximizes the log-likelihood under $M_0$. The
Schwarz criterion can be made to look more familiar by dividing the
$p_0 \times p_0$ matrix $\XX^T_0 \hat{\VV}{}^{-1}\XX_0$ by $n$ so that
after taking the determinant we obtain the additional term $\frac{1}{2}
p_0\log(n)$, and then writing $p_0=p-p_1$.

The Schwarz criterion (\ref{eqschwarz}) depends on the prior so, for
cases when informative priors are not available, it is useful to
consider using reference priors. Pauler presents Schwarz criteria using
unit-informa\-tion Gaussian and Cauchy reference priors. These criteria
depend on what she calls the effective sample size. Write $\be_0 =
(\beta_{01},\ldots, \beta_{0p_0})^T$ and $\XX_{i0}\be_0 = \XX_i^{(1)}\beta_{01} + \cdots+ \XX_i^{(p_0)}\beta_{0p_0}$. Then a fixed
effect parameter $\beta_{0k}$ has an \textit{associated random effect}
if its covariate vector $\XX_i^{(k)}$ is proportional to a column of
$\ZZ_i$ for $i = 1,\ldots,p_0$. The \textit{effective sample size} for
$\beta_{0k}$ is $E_k= m$ if $\beta_{0k}$ has an associated random
effect and $E_k = n$ otherwise. For the Gaussian prior
\[
S_{G} = \ell\bigl\{\hat{\bth}(\be_{00})\bigr\} - \ell(
\hat{\bth}) + \frac
{1}{2}\sum_{k=1}^{p_0}
\log(E_k),
\]
and for the Cauchy prior
\[
S_{C} = S_G + \log \bigl(\pi^{1/2}/
\bigl[2^{p_0/2}\Gamma\bigl\{(p_0+1)/2\bigr\}\bigr] \bigr).
\]
The effective sample size concept seems reasonable but it is important
to keep in mind that it is a result of the choice of prior which is
arbitrary and is\break not intrinsic to the problem. For example, for the
Gaussian prior, the variance is taken to be\break $\Delta^{1/2}(\XX^T_0 \hat
{\VV}^{-1}\XX_0)^{-1}\Delta^{1/2}$, where $\Delta=
\diag(E_1,\ldots,\break
E_{p_0})$. The log determinant of the variance is\break $ -(1/2) \log|\Delta|
+ (1/2)\log|\XX^T_0 \hat{\VV}{}^{-1}\XX_0|$ so, with this prior variance,
the log determinant term in~(\ref{eqschwarz}) is replaced by $(1/2)
\log|\Delta| = (1/2)\cdot  \sum_{k=1}^{p_0}\log(E_k)$.\break Other choices of
$\Delta$ would therefore lead to other criteria.

To explore the effective sample size concept further, consider the
random intercept model. Then $\ZZ= \bdiag(\one_{n_1}, \ldots, \one_{n_m})$ so
any fixed effect that is constant within clusters (i.e., a
cluster level covariate) has an associated random effect and any fixed
effect that varies within clusters does not. Suppose we have $a$
cluster level covariates. Then\break $\sum_{k=1}^{p_0} \log(E_k) = (p_0-a)\log
(n) + a\log(m) = \break p_0\log(n) + a\log(m/n)$ and this reduces to $p_0 \log
(n)$ if we have no cluster level covariate. Thus, if there is no
cluster level covariate, the Gaussian version of the Schwarz criterion
is the difference divided by $-2$ of two familiar terms of the form
\[
\BIC_{G} = - 2\ell(\hat{\bth}) + \log(n) p.
\]
The advantage of using $S_G$ rather than $\BIC_G$ is that it can be
applied to more general cluster models, but it has the disadvantage of
requiring us to compare pairs of explicit hypotheses. When using the
Schwarz criteria, it is a good idea to hold one of the hypotheses fixed
to simplify comparison (and computation); in the example given in her
paper, Pauler compares different models of interest to the null model
with only an intercept.

\citet{Jones2011} proposes using BIC with an alternative measure of the
effective sample size. In the linear regression model, the coefficient
of the intercept in the normal equations for the least squares
estimator is $n$; in the linear mixed model, the coefficient is $\one_n^T\VV^{-1}\one_n$.
Jones suggests that this coefficient be used as a
measure of sample size but, since it depends on the units of
measurement, $\VV^{-1}$ be replaced by the correlation matrix. If $\UU$
is the diagonal matrix with diagonal equal to the square root of the
terms on the diagonal of the $\VV$, the correlation matrix $\UU^{-1}\VV
\UU^{-1}$ is invariant to linear transformations of $\yy$. Jones'
measure of\vadjust{\goodbreak} effective sample size is then $\one_n^T\UU\VV^{-1}\UU\one_n$.
Jones gives expressions for some particular cases, noting that
when $\VV=\sigma^2\II$, the effective sample size reduces to $n$, for
the random intercept regression model $\sum_{i=1}^m n_i(\gamma^2 + \sigma^2)/(n_i\gamma^2+\sigma^2)$,
where $\gamma^2 = \Var(u_i)$, and the
longitudinal autoregressive model $\sum_{i=1}^m \{1 + (n_i-1)(1-\phi
)/(1+\phi)\}$. Both measures lie between $m$ and $n$, attaining these
bounding values as $ \gamma^2/(\gamma^2 + \sigma^2) \rightarrow1$ or
$\phi\rightarrow1$ (perfect correlation) and when $\gamma^2 = 0$ or
$\phi=0$ (zero correlation), respectively. In general, estimating the
parameters in $\UU$ and $\VV$ leads to the criterion
\[
\BIC_J = -2\ell(\hat{\bth}) + \log\bigl\{\one_n^T
\hat{\UU}\hat{\VV}{}^{-1}\hat {\UU}\one_n\bigr\}(p+q).
\]

We can compute a Bayes factor for comparing models with different
variance parameters but it is then difficult to obtain simple
approximations (like those given by \cite*{Pauler1998}) to the Bayes factor.
In particular, it is difficult to make subsets of the parameters in $\ta
$ null orthogonal and the boundary issues need to be taken into
account. Pauler, Wakefield and Kass (\citeyear{Pauler1999}) and Saville, Herring and Kaufman (\citeyear{Saville11}) ignore null
orthogonality but do acknowledge and try to deal with the boundary issues.

Pauler, Wakefield and Kass (\citeyear{Pauler1999}) approach the boundary issues in the variance
component model by assuming that the parameter space $\bTH$ can be
expanded to an open set $\bTH^o$ containing $\bTH$ so that the boundary
of $\bTH$ is interior to $\bTH^o$, applying the Laplace approximation
on $\bTH^o$ and then restricting it to $\bTH$. For selecting the
variance parameters $\ta$, partition $\ta=(\ta_0^T, \ta_1^T)^T$ into
$\ta_0\in R^{q_0}$, $\ta_1\in R^{q_1}$, $q=q_0+q_1$, and consider
testing the null hypothesis $M_0\dvtx \ta_0 = \zer$ against $M_1\dvtx \ta_0 \ne
\zer$. Using Laplace's method, Pauler, Wakefield and Kass (\citeyear{Pauler1999}) propose the approximation
\begin{eqnarray*}
\label{eqschwarz2} S &=& \ell\bigl\{\hat{\bth}{}^o(0)\bigr\} - \ell
\bigl(\hat{\bth}{}^o\bigr) - \tfrac{1}{2}q_0 \log (2
\pi) \\
&&{}+ \tfrac{1}{2} \log\bigl|\KK_{\ta_0|\ta_1}\bigl(\hat{\bth}{}^o
\bigr)\bigr| - \log\bigl\{ h(\hat{\ta}_0|\hat{\be}, \hat{
\ta}_1)\bigr\} \\
&&{}+ \log\bigl\{C_0^o/C_1^o
\bigr\} ,
\end{eqnarray*}
where $\hat{\bth}{}^o(0)$ maximizes the likelihood on $\bTH^o$ under
$M_0$, $\hat{\bth}{}^o$ maximizes the likelihood on $\bTH^o$, $\hat{\bth
}$ is the maximum likelihood estimate (i.e., maximizes the likelihood
on $\bTH$), $\KK_{\ta_0|\ta_1}(\bth) = \KK_{\ta_0\ta_0}(\bth) - \KK_{\ta
_0\ta_1}(\bth)\cdot \KK_{\ta_1\ta_1}(\bth)^{-1}\KK_{\ta_1\ta_0}(\bth)$ is
computed from the appropriate submatrices of the inverse of the
observed information matrix $\KK(\bth) = -\ell''(\bth)^{-1}$, $h(\ta_0|\be,\ta_1)$
is the conditional prior density under $M_1$ for $\ta_0$
given $\be$ and $\ta_1$, $C_0^o = \Pr[\mathcal{N}\{\hat{\ta}{}^o_1,
\KK_{0\ta_1\ta_1} (\hat{\bth}{}^o)\} \in\bTH]$ with $\KK_{0\ta_1\ta_1}$ the
submatrix of the inverse observed information matrix under $M_0$ for
$\ta_1$, and $C_1^o = \break \Pr[\mathcal{N}\{\hat{\ta}{}^o,   \KK_{\ta\ta}(\hat
{\bth}^o)\} \in\bTH]$\vadjust{\goodbreak} with $\KK_{\ta\ta}$ the submatrix of $\KK$ for
$\ta$. The quantities $C_0^o$ and $C_1^o$ are of the same form as
normalizing constants for truncated multivariate Gaussian densities.
Pauler, Wakefield and Kass (\citeyear{Pauler1999}) propose using a truncated Gaussian reference prior
which leads to
\begin{eqnarray*}
\label{eqschwarz2} S_{\mathit{TG}} &=& \ell\bigl\{\hat{\bth}{}^o(0)
\bigr\} - \ell\bigl(\hat{\bth}{}^o\bigr) + \tfrac
{1}{2}q_0
\log(n) \\
&&{}+ \log\bigl\{C_{\mathit{TG}}^oC_0^o/C_1^o
\bigr\},
\end{eqnarray*}
where $C_{\mathit{TG}}^o = \Pr[\mathcal{N}\{\zer,   n\KK_{\ta_0|\ta_1}(\hat{\bth
}{}^o)^{-1}\} \in\bTH]$ is the normalizing constant for the prior
density. Aside from the final boundary correction term, this is similar
to the usual Schwarz criterion. Under regularity conditions, the
boundary correction term is of smaller order than $\log(n)$ so, as
Pauler, Wakefield and Kass (\citeyear{Pauler1999}) note, the usual criterion can be used to select
variance parameters. In contrast to \citet{Pauler1998}, Pauler, Wakefield and Kass (\citeyear{Pauler1999})
do not attempt to make an adjustment for effective sample size.

Saville, Herring and Kaufman (\citeyear{Saville11}), following on from \citet{Saville09},
take a different approach to the boundary issue. They parametrize the
linear mixed model (\ref{eqremodel}) using the alternative Cholesky
factorization (\ref{eqmodchol}) so $\GA= \sigma\PH\GA^{\dagger}$,
where $\GA^{\dagger}$ is a lower triangular matrix with ones on the
diagonal and $\PH= \diag\{\exp(\phi_1), \ldots, \exp(\phi_s)\}$. The
matrix $\PH$ is $\DD^{\dagger}$ from (\ref{eqmodchol}) on the
logarithmic scale. Let $\ph= (\phi_1,\ldots,\phi_s)^T$ and let $\ga^{\dagger}$
be the vector of free parameters in $\ga^{\dagger}$. They
assume that $\sigma^{-2}$ has a gamma distribution and then integrate
both $\uu$ and $\sigma^2$ from the density of $\yy$ given $\be$, $\ph$,
$\ga^{\dagger}$, $\uu$ and $\sigma^2$ to obtain the density of $\yy$
given $\be$, $\ph$ and $\ga^{\dagger}$ which is a multivariate $t$
density. They then recommend adopting weakly informative priors for the
parameters and use Laplace approximations to approximate the Bayes
factor for comparing $M_0$ to $M_1$. They argue that the parameters in
the multivariate $t$ density do not have boundary constraints, but in
fact the boundary has been moved from zero to negative infinity and
this is not necessarily more convenient for computation.

\subsection{Other Criteria} \label{secother}

There are a number of criteria of a more or less arbitrary nature which
have been proposed for model selection. We describe some of these in
this section.

For the linear mixed model (\ref{eqremodel}) or the transformation
model (\ref{eqtrmodel}), \citet{Pu2006} suggest a Generalized
Information Criterion of the form
\[
\GIC_{\kappa_n} = -2 \ell(\hat{\bth}) + \kappa_n(p+q).
\]
This criterion combines both marginal AIC ($\kappa_n=2$) and BIC
($\kappa_n=\log(n)$) and allows greater flexibility in the choice of
$\kappa_n$. For example,\vadjust{\goodbreak} it includes the Hannan--Quinn (\citeyear{Hannan1979}) penalty
$\kappa_n=2\log\log(n)$ and the \citet{Bozdogan1987} penalty $\kappa_n= \log
(n)+1$, both of which are available in SAS \texttt{Proc Mixed}. Pu and
Niu also apply GIC with $\kappa_n= n^{1/2}$. For any choice of $\kappa_n$, Pu and Niu
suggest implementing GIC in two stages (first fix $\ta$
and select the model for $\be$ and then fix $\be$ and select the model
for $\ta$), but it is also possible to implement it directly. Pu and
Niu explore the asymptotic properties of the procedure for selecting
regression terms but not for selecting variance parameters.

The idea of treating $\be$ and $\ta$ separately and differently is
taken up by \citet{Jiang2003}. For any vector $\ba$, let $\Vert\ba
\Vert^2 = \ba^T\ba$. Then Jiang and Rao propose selecting the
regression parameter $\be$ using
\[
\bigl\Vert\bigl\{\II_n - \XX\bigl(\XX^T\XX
\bigr)^{-}\XX^T\bigr\}\yy\bigr\Vert^2 +
a_n p,
\]
where $a_n$ is a real, positive sequence satisfying some asymptotic
conditions and $\bA^{-}$ is a generalized inverse of $\bA$. 
Other than through the conditions on $a_n$, this criterion does not
depend on $\ta$ so this selection can be carried out separately. For
the variance component model, partition the set of matrices $\{\ZZ^{(1)},\ldots,
\ZZ^{(q_{\gamma})}\}$ into sets $L_k$ of matrices which
(together with $\XX$) span the same linear space so that the matrices
in $L_1$ have higher rank than those in $L_2$ and so on. Jiang and Rao
give the example of a 3-factor crossed design where $L_1$ contains the
3-way interaction, $L_2$ the 2-way interaction and $L_3$ the main
effects. Jiang and Rao suggest selecting the variance parameters $\ta$
sequentially, starting in $L_1$ and progressing through the remaining
sets of matrices. Let $\BB=(\XX, \ZZ)$ and $\BB_{-j}$ be $\BB$ omitting
$\ZZ^{(j)}$, $j \in L_1$. Then they select from $L_1$, the set of
indices $j$ for which, for any $1 < b <2$,
\begin{eqnarray*}
&&\frac{n-\rank(\BB)}{\rank(\BB) - \rank(\BB_{-j})} \\
&&\qquad{}\cdot \frac{\Vert\{
\BB(\BB^T\BB)^{-}\BB-\BB_{-j}(\BB_{-j}^T\BB_{-j})^{-}\BB_{-j}\}\yy\Vert^2}{\Vert\{\II_n -\BB(\BB^T\BB)^{-}\BB\}\yy\Vert^2}
\\
&&\quad > 1 + \bigl\{n-\rank(\BB)\bigr\}^{(b/2)-1} \\
&&\qquad{}+ \bigl\{\rank(\BB) - \rank(
\BB_{-j})\bigr\}^{(b/2)-1}.
\end{eqnarray*}
For the second group $L_2$, let $l_2$ denote a subset of indices in
$L_2$. Let $\BB_{1}(l_2) = (\XX,\ZZ^{(j)},j\in l_2\cup L_3,\cup L_4,
\ldots)$ be the matrix comprised of $\XX$ and the $\ZZ^{(j)}$, for $j$
from $l_2, L_3, L_4, \ldots.$ Then choose $l_2 \in L_2$ to minimize
\[
\bigl\Vert\bigl[\II_n - \BB_{1}(l_2)\bigl\{
\BB_{1}(l_2)^T\BB_{1}(l_2)
\bigr\}^{-}\BB_{1}(l_2)\bigr]\yy
\bigr\Vert^2 + a_{1n} \#(l_2),
\]
where $a_{1n}$ is a real, positive sequence satisfying some asymptotic
conditions and $\#(l_2)$ is the number of\vadjust{\goodbreak} parameters in $l_2$. Jiang
and Rao consider the penalties $a_{1n} \in\{2, \log(n), n/\log(n)\}$.
The procedure extends naturally to the remaining groups $L_3, L_4,
\ldots.$ Jiang and Rao give conditions under which the procedure is consistent.

\citet{Takeuchi1975} proposes using as a measure of model complexity $\trace
\{\KK(\bth)\LL(\bth)^{-1}\}$, where\break $\KK(\bth) = \Var\{\partial\ell
(\bth)/\partial\bth\}$ is the variance of the score function and $\LL
(\bth) = -\E\{\partial^2 \ell(\bth)/\partial\bth\,\partial\bth^T\}$ is
the expected information. As Burnham and Anderson\break [(\citeyear{Burnhama2002}), page~367]
note, this complexity measure can be expressed as $\trace\{\LL(\bth)\LL
(\bth)^{-1}\KK(\bth)\LL(\bth)^{-1}\}$,\break which is the trace of the
inverse of the asymptotic variance of $\hat{\bth}$ when the model holds
multiplied by the (sandwich) variance of $\hat{\bth}$ when the model
does not hold. If the model is correct, the measure reduces to $p+q$
and the Takeuchi Information Criterion
\[
\TIC = -2\ell(\hat{\bth}) + 2\trace\bigl\{\KK(\bth)\LL(\bth)^{-1}\bigr
\}
\]
is the same as $m\mathit{AIC}$. The Neural Information Criterion (NIC) of Murata,
Yoshizawa and Amari (\citeyear{Murata1994}) measures complexity in a similar way but uses the
regularized log-likelihood $\ell(\bth) + \log\{h(\bth)\}$ in place of
$\ell(\bth)$. Let $\KK_h(\bth) = \Var (\partial[\ell(\bth)+ \log\{
h(\bth)\}]/\partial\bth )$ and $\LL_h(\bth) = -\E (\partial^2
[\ell(\bth)+ \log\{h(\bth)\}]/\partial\bth\partial\bth^T )$. Then
the complexity measure in NIC, called the \textit{effective number of
parameters} by \citet{Moody1992}, is\break $\trace\{\KK_h(\bth)\cdot\LL_h(\bth)^{-1}\}
$. Ripley [(\citeyear{Ripley1996}), page~140] points out that the estimation of this
measure is generally not straightforward.

The minimum description length approach (MDL) developed by Rissanen in
the 1980s (see \cite*{Rissanen2007}) chooses the model that achieves maximum
data compression by minimizing the code length of the data and the
model. There are different coding schemes which lead to different MDL
criteria. The most relevant for the linear mixed model is the two-stage
code which leads to a penalized likelihood and is equivalent to $\BIC$,
the mixture scheme which produces a criterion that is related to a
Bayes factor and the normalized maximum likelihood scheme. For a
geostatistical model [the linear mixed model with $\GA=\zer$ and $\SI=
\sigma^2\RR(\de)$, where the parameters $\de$ describe the spatial
correlation between observations, so $\ta= (\de^T, \sigma^2)^T$ and
$q=q_{\delta}+1$], \citet{Hoeting2006} use the two-stage code and
propose the minimum description length criterion $\BIC/2$.
\citet{Liski2008} consider spline smoothing by fitting the random
effect model with one variance component ($q_{\gamma}=1$) and $\SI
=\sigma^2\II_n$. They use the normalized maximum likelihood\vadjust{\goodbreak} coding
scheme to produce the conditional criterion
\[
\MDL = - \ell(\hat{\bth}|\hat{\uu}) + \log\biggl[\int f\bigl\{\qqq|\hat{\uu}(\qqq);
\hat{\bth}(\qqq)\bigr\} \,d\qqq\biggr],
\]
where $f(\yy|\uu; \bth) = \exp\{\ell(\bth| \uu)\}$ is the conditional
density of $\yy|\uu$. The penalty term, called the \textit{parametric
complexity} of the model, is difficult to compute because the
conditional density is evaluated at the estimators before being integrated.

\citet{Kubokawa2011} introduces some prediction criteria which are variants
on Mallows $C_p$. Let $\tilde{\ta}$ be an estimator of $\ta$ from the
full model which satisfies a second order expansion like (\ref{eqtauexpansion}) $\tilde{\ta} - \ta= \tilde{\bt}_1(\ta) + \tilde{\bt}_2(\ta
) + O_p(n^{-3/2})$,
where $\E\{\tilde{\bt}_1(\ta)\} =
\E\{(\tilde{t}_{11}(\ta),\ldots,\break
\tilde{t}_{1q}(\ta))^T\} = \zer$, $\tilde{\bt}_1(\ta) = O_p(n^{-1/2})$
and $\tilde{\bt}_2(\ta) = \break(\tilde{t}_{21}(\ta), \ldots, \tilde
{t}_{2q}(\ta))^T = O_p(n^{-1})$. (He also considers estimating $\ta$
from the current candidate model but found that it performs poorly.)
Then let $\hat{\be}(\ta)$ be the generalized least squares estimator of
$\be$ defined in~(\ref{eqgls}), $\hat{\uu}(\ta) = \GA^{-1}\hat{\vv}(\ta
)$ be the BLUP of $\uu$ with $\hat{\vv}(\ta)$ defined in (\ref
{eqblup}), and let $\tilde{\PS}$, $\tilde{\SI}$ and $\tilde{\VV}$ be
estimators of $\PS$, $\SI$ and $\VV$ constructed using $\tilde{\ta}$.
Kubokawa defines
\begin{eqnarray*}
m\PEC &=& \bigl\{\yy- \XX\hat{\be}(\tilde{\ta})\bigr\}^T \tilde{
\VV}^{-1}\bigl\{\yy- \XX\hat{\be}(\tilde{\ta})\bigr\} \\
&&{}+ 2\bigl\{p +
q_m(\tilde{\ta})\bigr\} ,
\\
c\PEC &=& \bigl\{\yy- \XX\hat{\be}(\tilde{\ta}) - \ZZ\tilde{\PS}^{1/2}
\hat {\uu}(\tilde{\ta})\bigr\}^T \\
&&{}\cdot\tilde{\SI}^{-1}\bigl\{\yy-
\XX\hat{\be}(\tilde{\ta }) - \ZZ\tilde{\PS}^{1/2}\hat{\uu}(\tilde{\ta})
\bigr\}\\
&&{} + 2 \bigl\{\rho(\tilde{\ta }) + q_c(\tilde{\ta})\bigr\},
\end{eqnarray*}
where
\begin{eqnarray*}
&&q_m(\ta) \\
&&\quad= \frac{1}{2}\sum_{i=1}^q
\trace \biggl[\frac{\partial\VV
}{\partial\tau_i} \E \biggl\{ \frac{\partial^2 \tilde{t}_{1i}(\ta
)}{\partial\yy\,\partial\yy^T} \biggr\} \biggr] \\
&&\qquad{}+
\frac{1}{2} \sum_{i=1}^q \trace
\biggl(\frac{\partial\VV}{\partial\tau_i} \VV^{-1} \biggr)\E\bigl\{\tilde
{t}_{2i}(\ta)\bigr\}
\\
&&\qquad{} - \frac{1}{4}\sum_{i=1}^q\sum
_{j=1}^q \trace \biggl(\frac{\partial^2 \VV^{-1}}{\partial\tau_i\,\partial\tau_j}
\VV \biggr)\E\bigl\{\tilde{t}_{1i}(\ta )\tilde{t}_{1j}(\ta)
\bigr\} ,
\end{eqnarray*}
\begin{eqnarray*}
&&q_c(\ta)\\
&&\quad = -\frac{1}{2}\sum
_{i=1}^q \trace \biggl[\VV\frac{\partial(\VV^{-1}\SI\VV^{-1})}{\partial\tau_i}\VV\E
\biggl\{\frac{\partial^2 \tilde
{t}_{1i}(\ta)}{\partial\yy\,\partial\yy^T} \biggr\} \biggr]\\
&&\qquad{} + \frac{1}{2} \sum
_{i=1}^q \trace\biggl[\frac{\partial\SI}{\partial\tau_i}
\VV^{-1}\biggr]\E\bigl\{\tilde {t}_{2i}(\ta)\bigr\}
\\
&&\qquad{} - \sum_{i=1}^q\sum
_{j=1}^q \trace \biggl\{\frac{1}{4}
\SI^{-1}\frac
{\partial^2 \SI}{\partial\tau_i\,\partial\tau_j}\SI\VV^{-1}\\
&&\hspace*{95pt}{} - \frac
{\partial\SI}{\partial\tau_i}
\SI^{-1}\frac{\partial(\SI\VV^{-1})}{\partial\tau_j} \biggr\}\\
&&\hspace*{66pt}{}\cdot\E\bigl\{\tilde{t}_{1i}(
\ta)\tilde{t}_{1j}(\ta )\bigr\},
\end{eqnarray*}
and $\rho(\ta)$ is the effective degrees of freedom. The computations
are quite formidable.

Finally, \citet{Wu2002} and \citet{Fang2011} consider using
cross-validation to select linear mixed models. For the independent
cluster model with $\SI=\sigma^2\II_n$, the leave-one-cluster-out
criterion is
\begin{eqnarray*}
&& m^{-1}\sum_{i=1}^m
n_i^{-1}\bigl(\yy_i - \XX_i
\hat{\be}{}^{[i]}\bigr)^T\bigl(\ZZ_i\hat {
\PS}_{*}^{[i]}\ZZ_i^T+
\II_{n_i}\bigr)^{-1}\\
&&\quad{}\cdot\bigl(\yy_i -
\XX_i \hat{\be}{}^{[i]}\bigr),
\end{eqnarray*}
where $\hat{\be}{}^{[i]}$ and $\hat{\PS}{}^{[i]}_{*}$ are the maximum
likelihood estimators of $\be$ and $\PS_{*} = \PS/\sigma^2$ using the
data without cluster $i$; the leave-one-observation-out criterion is
\[
n^{-1}\sum_{i=1}^m \sum
_{j=1}^{n_i}\bigl(y_{ij} -
\xx_{ij}^T \hat{\be }{}^{[ij]} -
\zz_{ij}^T \hat{\vv}_i{}^{[ij]}
\bigr)^2,
\]
where $\xx_{ij}^T$ is the $j$th row of $\XX_i$, $\zz_{ij}^T$ is the
$j$th row of $\ZZ_i$, and $\hat{\be}{}^{[ij]}$ and $\hat{\vv}{}^{[ij]}_i$
are the maximum likelihood estimators and predictors of $\be$ and $\vv_i$, respectively, using the data without observation $j$ in cluster
$i$. The leave-one-cluster-out criterion is a marginal criterion,
whereas the leave-one-observation-out criterion is a conditional
criterion. \citet{Fang2011} shows that for $m \rightarrow\infty$ with
$n_i=n_1$ fixed (or $\bar{n} \rightarrow n_1$) and $\PS_{*}$ known, (i)
leave-one-cluster-out cross-validation and $m\mathit{AIC}$ of \citet{Vaida2005} are asymptotically equivalent, and (ii)
leave-one-observation-out cross-validation and $c\mathit{AIC}$ of \citet{Vaida2005} are asymptotically equivalent. This extends the
relationship between cross-validation and AIC in the linear regression
model established by \citet{Stone1977} to the linear mixed model.

\section{Shrinkage Methods} \label{seclasso}

One issue with the direct application of the information criteria
defined in Section~\ref{secinformation} is that they generally involve
comparing $2^{p+q}$ different models, which is not computationally
feasible when $p$ and/or $q$ is large. Even when $p+q \ll n$ is fixed, it
is still possible for $p+q$ to be large.
Shrinkage methods such as the LASSO (\cite{Tibshirani1996}) are popular for
selecting models in the linear regression setting when $p$ is of medium\vadjust{\goodbreak}
or large size due to its computational feasibility and statistical
accuracy (e.g., B\"uhlmann and van de Geer, \citeyear{Buhlmann2011}, page 20).
In this section we review the shrinkage approach to model selection in
the linear mixed model case.
We begin by discussing the linear regression case ($\SI= \sigma^2 \II_n$ and $\PS=\mathbf{0}$), since many of the ideas in the mixed
model case are motivated by this simpler case.

For the linear regression model, \citet{Tibshirani1996} proposes the LASSO
(least absolute shrinkage and selection operator)
method for simultaneous model estimation and selection. It is usual to
standardize the covariates $\XX$ and sometimes also to center $\yy$.
The selected model minimizes
%
\begin{equation}
\frac{1}{2}\Vert \yy- \XX\be\Vert ^2 + n \sum
_{j=1}^p\phi_{\lambda_j} \bigl( \mid
\beta_j \mid \bigr), \label{lasso}
\end{equation}
with respect to $\be=(\beta_1,\beta_2,\ldots,\beta_p)^T$,
where
%
\begin{eqnarray}
\phi_{\lambda_j}\bigl ( \mid\beta\mid \bigr)=\frac{\lambda_j\mid
\beta\mid}{2n} \quad\mbox{and}\quad
\lambda_j=\lambda,
\nonumber
\\[-8pt]
\\[-8pt]
\eqntext{ j=1,2, \ldots, p.} \label{lassopen}
\end{eqnarray}
When the tuning parameter $\lambda>0$ is large enough some of the
parameters in $\be$ are shrunk to exactly zero and, hence, minimizing
this criterion does model selection automatically. The minimization
problem (\ref{lasso}) with the LASSO penalty function (\ref{lassopen})
is a convex problem and there are efficient algorithms available to
compute the solution. For example, the LARS algorithm in \citet{Efron2004}
or the coordinate decent algorithms defined in \citet{Fried2007} and
Meier, van~de Geer and
B{\"u}hlmann (\citeyear{Meier2008}) can be applied.

There have been various further advances in penalized least squares
approaches for model selection since Tibshirani's original paper (e.g.,
see \cite*{Fan2010}, pages 107--117, and Tibshirani, \citeyear{Tibshirani2010}, for brief reviews).
One problem with the LASSO is that it tends to shrink large $\be$
coefficients too much, leading to bias issues (\cite{Fan2001}). As an
alternative to (\ref{lassopen}), \citet{Fan2001} suggest the
SCAD (smoothly clipped absolute deviation) penalty function defined by
its derivative
\begin{eqnarray*}
\phi^{\prime}_{\lambda_j}\bigl ( \mid\beta\mid \bigr)&=&\lambda_j
\biggl\{ I\bigl(\mid\beta\mid\leq\lambda_j\bigr) \\
&&\hspace*{18pt}{}+\frac{ ( a\lambda_j -\mid
\beta\mid )_{+}}{ ( a-1  )\lambda_j} I\bigl(\mid
\beta\mid > \lambda_j\bigr) \biggr\} \quad\mbox{and}\\
\lambda_j&=&
\lambda,\quad  j=1,2,\ldots, p,
\end{eqnarray*}
with $a=3.7$. They propose an algorithm based on local quadratic\vadjust{\goodbreak}
approximations and, more recently, \citet{Zou2008} propose a local
linear approximation, since the SCAD penalized loss function is
difficult to minimize directly due to the singularities in the penalty function.
\citet{Zou2006} introduces the ALASSO (Adaptive LASSO) which also helps
overcome the bias problems associated with the LASSO. The ALASSO
penalty function is
%
\begin{eqnarray}
\phi_{\lambda_j} \bigl( \mid\beta\mid \bigr)=\frac{\lambda_j\mid
\beta\mid}{2n} \quad\mbox{and}\quad
\lambda_j=\frac{\lambda}{\mid\hat
{\beta}_j \mid^{\iota}},
\nonumber
\\[-8pt]
\\[-8pt]
\eqntext{ j=1,2, \ldots, p,} \label{alasso}
\end{eqnarray}
where $\iota> 0$ is an additional parameter often taken to be equal to
1 and $\hat{\be}$ is a $n^{1/2}$-consistent estimator of~$\be$. \citet{Zou2006}
shows that the LARS algorithm can also be used to solve the
ALASSO minimization problem.

We now consider the linear mixed model case and assume $\PS$ has a
general form.
Bondell, Krishna and
Ghosh (\citeyear{Bondell2010}), \citet{Ibrahim2011} and \citet{Peng2012} are
to date the only authors to consider truly joint selection of both $\be
$ and $\ta$ using a shrinkage approach in the fixed parameter dimension
setting. Other authors apply shrinkage methods
to select on $\be$ only, assuming that the variance structure is not
subject to selection (e.g., Foster, Verbyla and
Pitchford, \citeyear{Foster2007}; Ni, Zhang and Zhang, \citeyear{Ni2010};
Wang, Eskridge and Crossa, \citeyear{Wang2010}).
We therefore focus on the methodology in Bondell, Krishna and
Ghosh (\citeyear{Bondell2010}), \citet{Ibrahim2011} and \citet{Peng2012}.
All three consider model selection for the independent cluster model
(\ref{eqremodelg}) assuming $\SI= \sigma^2 \II_n$ and both $s_i=s_1$
and $\PS_i=\PS_1$ are the same across clusters.
Both Bondell, Krishna and
Ghosh (\citeyear{Bondell2010}) and \citet{Ibrahim2011} use Cholesky
parametrizations and we will assume that $\GA_i$ is the Cholesky factor
of $\PS_i$ for the rest of this section.
Note that \citet{Ibrahim2011}
consider the more general mixed effects model setting where $\yy_i$
given $\uu_i$ and $\XX_i$ belong to the exponential family, but for
comparative purposes we will restrict the discussion to the Gaussian
case only.

\citet{Ibrahim2011} propose maximizing a penalized marginal log-likelihood
%
\begin{equation}\qquad
\ell(\bth) -m \sum_{j=1}^p
\phi_{\lambda_{j}} \bigl( \mid\beta_j \mid \bigr) - m\sum
_{k=1}^{s_1}\phi_{\lambda_{p+k}} \bigl( \Vert
\ga_k \Vert  \bigr), \label{lmmpen}
\end{equation}
with respect to $\bth$, where $\ga_k$ contains the nonzero elements in
the $k$th row of $\GA_i$ and $\ell(\bth)$ is defined in~(\ref
{logliktheta}). Either the SCAD or ALASSO penalty functions are used
in (\ref{lmmpen}) and there\vadjust{\goodbreak} are two tuning constants which are defined by
\begin{eqnarray}
\lambda_j=\lambda^{(1)},\quad j=1,2,\ldots, p\quad \mbox{and}\quad
\lambda_{p+k}=\lambda^{(2)}\sqrt{k},
\nonumber\\
\eqntext{ k=1,2,\ldots,
s_1.}
\end{eqnarray}
The ALASSO penalty functions differ slightly from (\ref{alasso}) and
are defined as
\begin{eqnarray*}
\phi_{\lambda_j} \bigl( \mid\beta\mid \bigr)&=&\lambda_j\frac{\mid
\beta\mid}{\mid\hat{\beta}_j\mid},\quad
j=1,2, \ldots, p \quad\mbox {and}\\
 \phi_{\lambda_{p+k}} \bigl( \Vert \ga_k
\Vert  \bigr)&=&\lambda_{p+k}\frac{ \Vert \ga_k \Vert }{ \Vert \hat{\ga}_k
\Vert },\quad k=1,2, \ldots,
s_1,
\end{eqnarray*}
where $\hat{\be}$ and $\hat{\ga}_k$ are the unpenalized maximum
likeli\-hood estimators. Notice that the parameters $\ga$ are~se\-lected in
a grouped manner similar to the LASSO~for grouped variables (\cite{Yuan2006}), and this helps preserve the positive definite constraint in
$\PS$.

Bondell, Krishna and
Ghosh (\citeyear{Bondell2010}) use the alternative Cholesky factor parametrization
$\GA_i=\sigma\DD_i^{\dagger}\GA_i^{\dagger}$,
where $\DD_i^{\dagger}=\operatorname{diag}(d_1,d_2,\ldots, d_{s_1})^T$ is a
diagonal matrix and $\GA_i^{\dagger}$, whose $(l,r)$th element is
$\gamma^{\dagger}_{lr}$, is a $s_1 \times s_1$ lower triangular matrix
with ones on the diagonal. Setting $d_l=0$ is equivalent to setting all
the elements in the $l$th column and $l$th row to zero and, hence, a
single parameter controls the inclusion/exclusion of a group of random
effects. Let $\dd=(d_1,d_2,\ldots, d_{s_1})^T$, let $\ga^{\dagger}$ be
the vector of free parameters in $\GA_i^{\dagger}$ and define $\bth^{\dagger}=(\be^T,\dd^T,\ga^{\dagger T})^T$. Note that $\sigma^2$ is
not included in $\bth^{\dagger}$. Bondell, Krishna and
Ghosh (\citeyear{Bondell2010}) propose
maximizing an ALASSO penalized log-likelihood
\[
\ell\bigl(\bth^{\dagger}\bigr) -\lambda^{(3)} \Biggl( \sum
_{j=1}^p \frac{\mid\beta_j \mid}{\mid\hat{\beta}_j \mid} + \sum
_{k=1}^{s_1} \frac{\mid d_k \mid
}{\mid\hat{d}_k \mid} \Biggr)
\]
with respect to $\bth^{\dagger}$, where $\lambda^{(3)}$ is a single
tuning constant. Here $\hat{\be}_j$ are the unpenalized generalized
least squares estimates and the $\hat{d}_k$ is obtained from
decomposing the unpenalized restricted maximum likelihood estimate of
$\PS_i$.

The Cholesky decompositions prove to be very helpful in estimation. The
conditional expectations of $\yy$ given $\uu$ can be rearranged to give
%
\begin{equation}\label{linearIb}\qquad
\XX\be+\ZZ\GA\uu= %
\bigl(\matrix{ \XX& \bigl( \uu^T \otimes
\ZZ \bigr) \JJ_{ms_1} }\bigr) %
\pmatrix{ \be\vspace*{2pt}
\cr
\ga }
,
\end{equation}
where $\operatorname{Vec}(\GA)=\JJ_{ms_1}\ga$ [the matrix $\JJ_{ms_1}$
transforms $\ga$ to $\operatorname{Vec}(\GA)$], or
%
\begin{eqnarray}\label{linearBo}\qquad
&&\XX\be+\ZZ\GA\uu
\nonumber
\\[-8pt]
\\[-8pt]
\nonumber
&&\quad= %
\bigl(\matrix{ \XX& \ZZ\operatorname{diag} \bigl(
\GA^{\dagger} ( \sigma\uu ) \bigr) ( \mathbf{1}_m \otimes
\II_{s_1} ) }\bigr) %
\pmatrix{ \be\vspace*{2pt}
\cr
\dd }
.
\end{eqnarray}
The conditional expectations can therefore be written in a form which
is linear in the parameters that are subject to selection.
Bondell, Krishna and
Ghosh (\citeyear{Bondell2010}) and \citet{Ibrahim2011} both adapt the EM
algorithm to estimate the parameters. They treat $\uu$ as unobserved in
the E-step and the M-step involves maximizing a penalized objective
function. To incorporate grouped penalization, \citet{Ibrahim2011}
use a modification of the local linear approximation algorithm proposed
by \citet{Zou2008}. Bondell, Krishna and
Ghosh (\citeyear{Bondell2010}) in their M-Step apply a
standard quadratic programming technique. The EM penalized maximum
likelihood estimators above are obtained first by assuming $\uu$ is
known, then $\GA$ is estimated and then $\uu$ estimated. This process
differs subtly from the information criteria approaches in Section \ref
{secinformation}, where a different order is used when deriving the
criteria there: first it is assumed that $\GA$ is known, then $\uu$ is
estimated and then $\GA$ is estimated.

Although the approaches of Bondell, Krishna and
Ghosh (\citeyear{Bondell2010}) and \citet{Ibrahim2011} share some elements in common, there are some differences
between them which are important to highlight. Bondell, Krishna and
Ghosh (\citeyear{Bondell2010})
incorporate a single tuning constant which is the same for penalizing
both $\be$ and $\ta$, whereas \citet{Ibrahim2011} have a more
flexible approach with
two different tuning constants. Bondell, Krishna and
Ghosh (\citeyear{Bondell2010}) use $\dd$ rather
than $\ga$ in model selection and they effectively treat $\ga^{\dagger
}$ like nuisance parameters since they do not appear in the penalty.
Neither Bondell, Krishna and
Ghosh (\citeyear{Bondell2010}) nor \citet{Ibrahim2011} incorporate
$\hat{\uu}$ into the penalized likelihood criterion and their methods
are therefore more in line with the marginal information criteria of
Section~\ref{marginalAIC}, rather than the conditional approach of
Section~\ref{condAIC}.

One open issue with both Bondell, Krishna and
Ghosh (\citeyear{Bondell2010}) and \citet{Ibrahim2011} is that the Cholesky decompositions are dependent on the order
in which the random effects appear and are not permutation invariant
(\cite{Pou2011}). This means in the finite sample case that different
model selections result from using different orders in the columns of
$\ZZ_i$.
We confirmed this by running the first simulation example in Bondell, Krishna and
Ghosh (\citeyear{Bondell2010}) with different orders in the columns of $\ZZ_i$. Note also
that setting $d_k$ and $\ga_k$ to zero is not equivalent to setting the
$k$th diagonal element in $\PS_i$ to zero, which for the independent
cluster model (\ref{eqremodelg}) is the more natural selection problem.
Another issue is that both Bondell, Krishna and
Ghosh (\citeyear{Bondell2010}) and \citet{Ibrahim2011}\vadjust{\goodbreak} use the unpenalized maximum likelihood or restricted maximum
likelihood estimates as the weights in the ALASSO penalty, but in
practice unpenalized maximum likelihood algorithms often fail to
converge when the underlying $\ta$ is sparse and/or $p$ is large (e.g.,
\cite*{Ng2012}, page~310; Jiang, Luan and Wang, \citeyear{Jiang2007b}, page 2252). Also,
some of the maximum likelihood estimates of variance parameters could
be exactly on the zero boundary, implying that the ALASSO weight is
infinity. Note that boundary problems do not occur in the regression
case since only $\be$ is penalized.

\citet{Peng2012} also apply a shrinkage method, although their
approach is quite different from Bondell, Krishna and
Ghosh (\citeyear{Bondell2010}) and \citet{Ibrahim2011}. Instead of doing selection on $\PS_1$ directly,
\citet{Peng2012} select the random effects by penalizing $\vv=\GA\uu$. Write $\PS_1=\sigma^2\PS_1^{\dagger}$
and then, motivated by an asymptotic
expansion, estimate $\PS_1^{\dagger}$ by
%
\begin{equation}
\hat{\PS}_1^{\dagger}=\frac{\sum_{i=1}^m\vv_i \vv_i^T}{m\hat{\sigma
}{}^2}-\frac{\sum_{i=1}^m (\ZZ_i^T\ZZ_i)^{-1}}{m}.
\label{psupdate}
\end{equation}
To estimate and select the model, \citet{Peng2012} define the
following simple iterative procedure which penalizes both $\be$ and $\vv$:
\begin{longlist}[(1)]
\item[(1)] For each $i$ update $\vv_i$ given $\be$ by minimizing with
respect to $\vv_i$ the penalized least squares criterion
\begin{eqnarray*}
&&(\yy_i -\XX_i\be-\ZZ_i\vv_i
)^T (\yy_i -\XX_i\be-\ZZ_i
\vv_i )\\
&&\quad{}+ 2n\sum_{k=1}^{s_1}
\phi_{\lambda^{(4)}}\Bigl(\sqrt{\bigl\mid\hat{\ps }_{kk}^{\dagger}
\bigr\mid}\Bigr),
\end{eqnarray*}
where $\hat{\ps}{}^{\dagger}_{kk}$ is the $k$th diagonal element of $\hat
{\PS}_1^{\dagger}$. Then update $\PS_1^{\dagger}$ using (\ref{psupdate}).
\item[(2)] Update $\be$ given $\PS_1^{\dagger}$ by minimizing with
respect to $\be$ the penalized least squares criterion
\begin{eqnarray*}
&&( \yy-\XX\be )^T \bigl( \II_n+ \ZZ\PS^{\dagger}
\ZZ^T \bigr)^{-1} ( \yy-\XX\be )\\
&&\quad{}+ 2n\sum
_{k=1}^p \phi_{\lambda
^{(5)}}\bigl(\mid
\beta_k \mid\bigr),
\end{eqnarray*}
where $\PS^{\dagger}=\operatorname{block diag}  (\PS^{\dagger}_1,\PS^{\dagger}_1,\ldots, \PS^{\dagger}_1  )$ has $m$ identical blocks
on the diagonal.
\end{longlist}
In both cases the SCAD penalty function is used with tuning constants
$\lambda^{(4)}$ and $\lambda^{(5)}$.

One advantage of the \citet{Peng2012} selection method is that the
random effects $\vv$ are unconstrained and are treated like unknown
regression coefficients, which make the selection and computations easy\vadjust{\goodbreak}
to handle. In comparison, the optimization procedures in both Bondell, Krishna and
Ghosh (\citeyear{Bondell2010}) and \citet{Ibrahim2011} are slow and complex and can
sometimes fail to converge, especially when the underlying covariance
matrices are sparse and the tuning constants are small.
Another advantage of the \citet{Peng2012} approach is that it is
permutation invariant and does not depend on the order in which the
random effects appear. However, the estimate of $\PS_1^{\dagger}$ is
not always guaranteed to be positive semidefinite and further
adjustments may be needed (\cite*{Peng2012}, page 114).

Some further insight is obtained by comparing the asymptotic results in
Bondell, Krishna and
Ghosh (\citeyear{Bondell2010}), \citet{Ibrahim2011} and \citet{Peng2012}. In
the linear regression setting
\citet{Zou2006} proves that the ALASSO estimators possess oracle properties
asymptotically. That is, as $n\rightarrow\infty$ with $p < \infty$
fixed they (a) identify the true model and (b) achieve the optimal
estimation rate (i.e., the estimator performs as well as if the true
model were known in advance).
Similarly, Bondell, Krishna and
Ghosh (\citeyear{Bondell2010}) show that their penalized maximum
likelihood estimators possess the oracle property under some regularity
conditions and
\[
m \rightarrow\infty,\quad \lambda^{(3)} \rightarrow\infty \quad\mbox {and}\quad
\frac{\lambda^{(3)}}{\sqrt{m}}\rightarrow0
\]
with finite cluster sizes
$1 \leq n_i \leq K$, for some $K < \infty$ and $i=1,2,\ldots, m$.
\citet{Ibrahim2011} also prove that their procedure has the oracle
property under some regularity conditions. Let ${\be}_{t}$ and ${\ga
}_{k,t}$ be the true values of $\be$ and $\ga_k$, $ k=1,2,\ldots, s_1$,
respectively. Define
\begin{eqnarray*}
&&b_m=\min \Bigl[ \min_{j=1,\ldots,p}{\{\lambda_j\dvtx {
\be}_t=0 \}}, \\
&&\hspace*{49pt}\min_{k=1,\ldots,s_1}{\bigl\{\lambda_{p+k}\dvtx \Vert {\ga}_{k,t} \Vert =0 \bigr\}} \Bigr]
\end{eqnarray*}
and
\begin{eqnarray*}
&&c_m=\max \Bigl[ \max_{j=1,\ldots,p}{\{\lambda_j\dvtx {
\be}_t \neq0 \}},\\
&&\hspace*{49pt} \max_{k=1,\ldots,s_1}\bigl\{\lambda_{p+k}\dvtx \Vert {\ga}_{k,t} \Vert \neq0 \bigr\}  \Bigr].
\end{eqnarray*}
The limit conditions are
\[
m \rightarrow\infty,\quad \sqrt{m}b_m \rightarrow\infty \quad\mbox {and}\quad
c_m \rightarrow0.
\]
\citet{Peng2012} show that their method is a consistent variable
selection procedure with some oracle properties, but the extra
condition $s_1 < m^{-1}\sum_{i=1}^mn_i$ is needed. As noted by \citet{Peng2012},
when the cluster sizes are small their method does not
perform as well (and is not as efficient) as methods based on the
marginal distribution. Note that\vadjust{\goodbreak} both Bondell, Krishna and
Ghosh (\citeyear{Bondell2010}) and \citet{Ibrahim2011} use the marginal distribution when deriving their
shrinkage estimators, which is an advantage in this case.

The shrinkage methods discussed above produce estimates of the model
parameters and select a model conditional on the tuning constants being
known. By varying the values of the tuning constants from large to
small, a path through the model space is defined where more parameters
get selected as $\lambda^{(1)}$, $\lambda^{(2)}$, $\lambda^{(3)}$,
$\lambda^{(4)}$ and $\lambda^{(5)}$ each approach zero. Model selection
on the path is reduced to selecting the values of the tuning constants.
This is one of the major advantages of shrinkage methods over direct
application of information criteria: shrinkage methods do not need to
consider all possible models (which is often not computationally
feasible when $p$ and $s_1$ are large), but only the models identified
on the path. Once the path is identified,
information criteria, cross-validation or other methods can then be
used to select the model from the path (see Section~\ref{secinformation}
for further details). The Fence method described in
Section~\ref{secfence} also uses a similar concept where models within
a ``fence'' are first identified, and then the second step chooses the
least complex model.

The choice of tuning constant is important because this ultimately
controls which model gets selected. Bondell, Krishna and
Ghosh (\citeyear{Bondell2010}) choose the
tuning constant to minimize the BIC type criterion
%
\begin{equation}
-2\ell\bigl(\hat{\bth}{}^{\dagger}\bigr) + \log(n)\#\bigl(\hat{
\bth}{}^{\dagger}\bigr), \label{tune1}
\end{equation}
over a grid of $\lambda^{(3)}$ values, where $\#(\hat{\bth}{}^{\dagger})$
is the number of nonzero elements in $\hat{\bth}{}^{\dagger}$.
\citet{Ibrahim2011} consider the broader class of generalized linear
mixed models where often the marginal likelihood is not directly
available. However, in the case of the linear mixed model, the marginal
likelihood is available and \citet{Ibrahim2011} would apply the BIC criterion
%
\begin{equation}
-2\ell(\hat{\bth}) + \log(m)\#(\hat{\bth}) \label{tune2}
\end{equation}
directly. There are clearly differences between (\ref{tune1}) and~(\ref
{tune2}). The $\bth^{\dagger}$ in Bondell, Krishna and
Ghosh (\citeyear{Bondell2010}) does not
include $\sigma^2$, whereas \citet{Ibrahim2011} do include $\sigma^2$
and so $\#(\hat{\bth}{}^{\dagger})$ and $\#(\hat{\bth})$ are slightly different.
In the linear mixed model, the definition of the effective sample size
is not obvious and has long been an issue for debate.
Bondell, Krishna and
Ghosh (\citeyear{Bondell2010}) use the total sample size $n$ in (\ref{tune1}),
but (\ref{tune2}) uses the total number of clusters~$m$. Another
alternative is to estimate the effective sample size by incorporating
an\vadjust{\goodbreak} estimate of the correlation matrix as suggested by \citet{Jones2011} (see
$\BIC_J$ in Section~\ref{secbic}), which leads to an estimate of the
effective sample size between $m$ and~$n$. A referee pointed out that
using information criteria to choose the tuning constants here has not
been rigorously justified and is somewhat ad-hoc. The issue is that the
number of nonzero estimated parameters corresponding to a given tuning
constant is not the same as the fixed number of independent parameters
under an assumed model.

An alternative way of choosing the tuning constant is to treat it like
an additional variance component in the model to be estimated directly
along with $\ta$. A similar approach is often used in the~semi\-parametric regression literature when estimating tuning constants
associated with penalized splines (Ruppert, Wand and Carroll, \citeyear{Ruppert22003}, page 108).
\citet{Tibshirani1996} notes that
$ |\beta_j|$ is proportional to (minus) the log density of the double
exponential distribution.
Foster, Verbyla and
Pitchford (\citeyear{Foster2007}) incorporate a LASSO penalty for $\be$ into a
linear mixed model and for estimation of the tuning constant each $\beta_k$ is
assumed to have a double exponential distribution with variance
${2}/{\lambda^2}$, where $\lambda$ is the tuning constant (so $\lambda$
is effectively treated like a hyperparameter in a hierarchical model).
Estimation of $\lambda$ is then carried out by maximizing an
approximate marginal log-likelihood. \citet{Ibrahim2011} also use a
similar idea for estimating their two tuning constants $\lambda^{(1)}$
and $\lambda^{(2)}$, however, they note that the estimates produced
from this method lead to significant overfitting.

\section{Fence Methods} \label{secfence}

Alternative model selection methods to information criteria or
shrinkage methods are rare and typically ad-hoc. A notable exception is
the Fence meth\-od for selecting predictors for complex models, which was
recently proposed by \citet{Jiang2008}. The Fence method is
computationally very demanding, particularly because it involves the
estimation of the standard deviation of the difference of lack-of-fit
measures, for example, the negative log-likelihood as in Section \ref
{secinformation}, the residual sum of squares or any appropriate
estimated loss, denoted by $Q_M = Q_M(\bth_M)$, $M\in\mathcal{M}$,
satisfying $Q_{M_2} \leq Q_{M_1}$ if $M_1 \subset M_2$. For example,
$Q_M = [\yy- \E_M (\yy)]^T [\yy- \E_M (\yy)]$. The Fence procedure in
\citet{Jiang2008} requires the calculation of
\[
\hat\sigma_{M,\tilde M}= \sqrt{\hat{\Var}\bigl[Q_{M}(
\bth_{M}) -Q_{\tilde
M}(\bth_{\tilde M})\bigr]}
\]
for all models $M\in\mathcal{M}$, where $\tilde M$ has the smallest
loss among all considered models. Jiang, Nguyen and Rao (\citeyear{Jiang2009}) reduce to some
extent the computational burden of the Fence method in their Simplified
Adaptive Fence procedure, which can be very competitive in
lower-dimensional problems and\break where convergence of estimation
procedures is not of a concern, such as when using the least squares
estimator in linear regression with $\XX^T\XX$ of full rank.

The key idea behind the Fence method is to estimate the loss for any
correct model $M_l$ by $Q_{M_l}({\hat\bth_{M_l}})$, which satisfies a
range of regularity conditions and is used to construct a fence. In
practice, $M_l$ can be the full or any other sufficiently large model.
The first step is to identify models $M\in\mathcal{M}$ inside the
fence, that is, models satisfying
%
\begin{equation}
Q_{M} \leq{Q}_{M_l} + b_n \hat
\sigma_{M,M_l}, \label{eqnfence1}
\end{equation}
where $b_n$ is a sequence of tuning constants.

\begin{figure*}[b]

\includegraphics{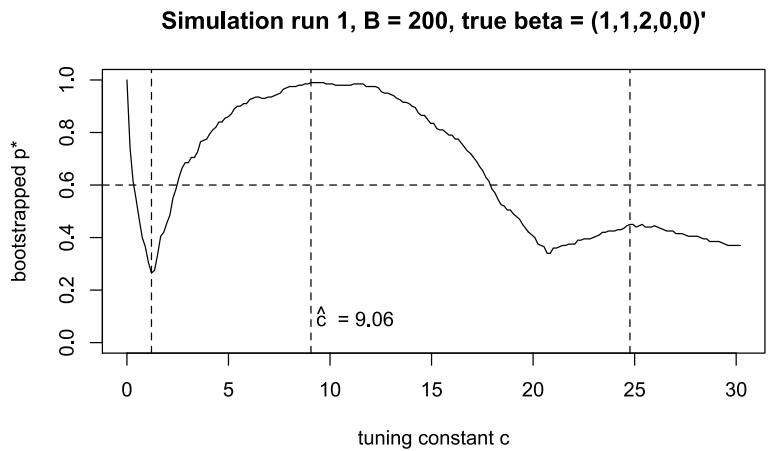}

\caption{A plot of $p^*$ based on the first simulated data set under
the simple linear mixed model $y_{ij} = \beta_0+\beta_1 x_{i1}+\beta_2x_{i1} +
\gamma u_i + \sigma\varepsilon_{ij},$ $i=1,\ldots,10 = m$,
$j=1,\ldots,5$, $\gamma=\sigma=1$ and $u_i,\varepsilon_{ij}\sim\mathrm{independent}\ \mathcal{N}(0,1)$.}
\label{fig-fence1}
\end{figure*}

The second step of Fence is to identify the least complex model within
the fence. If there is more than one such candidate, the model with the
smallest lack-of-fit measure is selected.
Conceptually,\break Fence shares a major advantage with shrinkage methods
(see Section~\ref{seclasso});
they both consider only a small proportion of models in $\mathcal{M}$,
although they choose the subset differently and select from the subset
differently. For Fence, only a small number of models satisfy (\ref
{eqnfence1}) when $b_n$ is small and these models can be identified
economically through backward or forward search algorithms.
The calculation of $Q_M(\hat{\bth}_M)$ is often straightforward,
particularly when $Q_M$ is the negative log-likelihood or residual sum
of squares. Using the residual sum of squares can be promising when
focus is on the selection of regression parameters that relate to the
whole population, but it could be more natural to use the conditional
log-likelihood when the selection focuses mainly on parameters
describing clusters.

The Simplified Adaptive Fence procedure, a computationally simpler
version to Adaptive Fence introduced in \citet{Jiang2008}, absorbs
the difficult quantity $\hat\sigma_{M,M_l}$ and the tuning constant
$b_n$ into a single constant,
%
\begin{equation}
Q_{M} \leq{Q}_{M_l} + c_n. \label{eqnfence2}
\end{equation}
Thus, the model selection problem turns into optimally choosing the
tuning constant $c_n$. Jiang,\break Nguyen and Rao (\citeyear{Jiang2009}) suggest calculating for
each\break $M\in\mathcal{M}$ the bootstrapped probability $p^*(M;c_n) =\break \P^*(M_0(c_n)=M)$,
where $M_0(c_n)$ is the optimal model satisfying (\ref
{eqnfence2}).
Jiang, Nguyen and Rao (\citeyear{Jiang2009})\vadjust{\goodbreak} calculate $p^*(M;c_n)$ with a parametric bootstrap
under $M_a$, a large correct model with at least one redundant
component. $M_a$ can be the full or any large model which is known to
be correct but not optimal. On the other hand, if the full model might
be the optimal model, $\XX$ can be extended to $(\XX,\xx_a)$. In our
own simulations we used $\xx_{ia} = (\frac{1}{p}\sum\xx_j)_{l_i}$,
where $(l_1,\ldots,l_n)$ is a random permutation of $\{1,\ldots,n\}$.
Adding this additional explanatory variable worked well in our
simulations, but there are many other possibilities. Jiang, Nguyen and Rao (\citeyear{Jiang2008,Jiang2009})
give an elaborate explanation of why such an adjustment is
required. Essentially, it ensures that the function $p^*(c_n) = \max_M
p^*(M;c_n)$ has desirable theoretical features. In particular, the
model that corresponds to the first significant peak at $\hat{c}_n$,
that is, $\hat{M}(\hat c_n) = \argmax_M p^*(M;\hat c_n)$, is a
consistent estimate of a correct model $M_l$ satisfying $M_a \supset
M_l \supseteq M_t$, provided the true model $M_t$ exists and the true
model is not the model used for the generation of the parametric
bootstrap samples, that is, \mbox{$M_t \neq M_a$}.
Jiang, Nguyen and Rao (\citeyear{Jiang2009}) state a theorem, which (under some technical
regularity conditions) establishes the existence of a $\tilde c$
(depending on $n$), which is at least a local maximum and an
approximate global maximum of $p^*(\tilde c)$, such that the
corresponding $\hat M(\tilde c)$ is consistent---in the sense that for
any $\kappa_1,\kappa_2 > 0$, there exist $n_{\operatorname{min}}$ and $B_{\operatorname
{min}}$ such that
%
\begin{eqnarray}\label{eqnproof-of-adafence}\quad
\P\bigl(p^*(\tilde c) \geq1 - \kappa_1\bigr) \wedge\P\bigl(\hat M(
\tilde c) = M_t\bigr) \geq1-\kappa_2
\nonumber
\\[-8pt]
\\[-8pt]
\eqntext{\mbox{if } n \geq
n_{\operatorname{min}} \mbox{ and } B \geq B_{\operatorname{min}}.}
\end{eqnarray}
Jiang, Nguyen and Rao (\citeyear{Jiang2009}) refer for the proof of (\ref
{eqnproof-of-adafence}) to the proof of Theorem 3 in \citet{Jiang2008}. For specific choices of $Q_M$ and $\mathcal M$ it could require
some care to show that all the regularity conditions hold. Empirically,
we confirmed that the first significant peak, which occurs at $\hat
c_n$, satisfies\vspace*{1pt} $p^*(\hat M; \hat c_n)\approx1$ for $\hat M \neq M_a$,
where $\hat c_n$ is the smallest possible choice of the tuning constant
when the peak is a plateau. For small to moderate $n$ and for larger
true models the peaks relating to the true model tend to be smaller
than one, whereas for small true models we often observed $p^*(M_t;
c_n) = 1$ for $c_l< c_n< c_u$. In our own simulations we found that the
following rule establishes a surprisingly successful and ``simple''
estimator of the true model: Consider only $c$ values with $p^*(c_n)$
attained by some $\hat M \subset M_a$; choose the first $\hat c_n$,
which is either a peak larger than some arbitrary value $\tau$ in
$(0.5,1)$ or the smallest $c_n$ value having maximal $p^*(c_n)$ value.
In our own implementations we used $\tau= 0.6$, which was chosen
before running any simulations, by a visual inspection of all published
results in the series of Fence papers.
(Jiang, Nguyen and Rao, \citeyear{Jiang2009}, suggest another adjustment, based on lower bounds
of large sample 95\% confidence intervals, which depend on the
bootstrap sample size and $p^*$.)

Figure~\ref{fig-fence1} shows a plot of $p^*$ over an appropriate range
of the tuning constant $c_n$. The data generating model is a $m=10$
independent cluster model with group sample sizes $n_i \equiv5$. The
full model has four covariates and an intercept, and the true model has
parameter vector $\be_t = (1,1,2,0,0)^T$. Responses were generated by
$y_{ij} = \xx^T_i\be+ \gamma u_i + \sigma\varepsilon_{ij}$, $i=1,\ldots
,\break 10 = m$, $j=1,\ldots,5$, $\gamma=\sigma=1$ and $u_i,\varepsilon_{ij}\sim
\mathrm{indepen}\mbox{-}\break \mathrm{dent}\ \mathcal{N}(0,1)$ with $x_{i1} = 1$ and the
remaining explanatory variables generated independently from $\mathcal
{U}(-2,2)$. We used the \texttt{lme()} function of the \texttt{nlme}
R-library to fit a total of $2^4+1 = 17$ linear mixed models as
described above having the same variance parameters. The Simplified
Adaptive Fence procedure with $Q(\bth) = -2\ell(\bth)$ correctly
estimates the true model and the corresponding peak occurs at $\hat
{c}_n = 9.06$. Three additional models have peaks: the too large model
used for the generation of the bootstrap samples at $c=0$, the correct
model having only $\beta_4 = 0$ at $c=1.21$ (which is a local maximum
difficult to detect by visual inspection) and the incorrect model with
$\be=(\beta_0,0,\beta_2,0,0)^T$ with a peak $p^*(24.77) = 0.450$.

A major attraction of the Simplified Adaptive\break Fence is its generality.
On the other hand, since the Simplified Adaptive Fence is heavily based
on bootstrapping from a too large correct model, it highlights any
computational limitations in the available estimation procedures.
In our simulations we noticed that fitting linear mixed models with
redundant random effects can be problematic. For example, we repeatedly
generated data from the same data generating model as in Bondell, Krishna and
Ghosh (\citeyear{Bondell2010}), Example 1. Using \texttt{lme()} and maximum likelihood, we
found that in seven out of the first ten simulation runs the estimates
failed to converge. The function \texttt{lmer()} from the R-package
\texttt{lme4} never failed in the first thousand simulation runs but
produced seven warnings of the type \texttt{In mer$\_$}\break\texttt{finalize(ans):
singular convergence (7)} and, more severely, calculating an auxiliary
quantity such as $\hat\PS^{-1}$ failed in five out of the first ten
simulation runs returning the warning \texttt{Error in
solve.}\break\texttt{default(VarCorr(M)\$} \texttt{grp): system is compu-}\break\texttt{tationally singular:}
\texttt{reciprocal condition}\break \texttt{number}.  This is in fact a problem for
most methods, including information criteria and shrinkage methods.

We conclude that using the Simplified Adaptive Fence can be attractive
when convergence is not a concern. However, it is potentially tedious
to implement the Simplified Adaptive Fence in simulation studies that
automatically loop through many runs of fitting mixed models with
redundant random terms. This is a possible explanation for why Jiang, Nguyen and Rao (\citeyear{Jiang2008,Jiang2009}) focused in their simulations on the selection of $\be$
only, and demonstrated that the Simplified Adaptive Fence can
successfully deal with linear mixed models as long as interest focuses
on selecting the regression parameters.

Recent work on the Invisible Fence (Jiang, Nguyen and Rao, \citeyear{Jiang2011}) and the
Restricted Fence\vadjust{\goodbreak} (\cite{Ng2012}) explores some ways to reduce
the computational burden. Just like the Fence and Simplified Adaptive
Fence, the Invisible Fence is based on the principle of selecting the
model within the fence that has minimum dimension and minimum $Q_M$
among other models within the fence of the same dimension. Jiang, Nguyen and Rao
(\citeyear{Jiang2011})\break showed that the model selected by the Simplified Adaptive Fence
is one of the models that minimizes $Q_M$ at each model dimension. This
means that if we can find this small set of models (one for each model
dimension), the model selection problem is considerably simplified. The
Invisible Fence uses the bootstrap to find the reduced set of models.
Specifically, for the $b$th bootstrap sample, for each model dimension
$j$, find the model $M_{bj}^*$ of dimension $j$ that minimizes
$Q_{Mb}^*$, the loss $Q_M$ computed for the $b$th bootstrap sample.
Then, for each fixed model dimension $j$, find the most frequently
selected model across bootstrap samples $M_j^*$ and its bootstrap
selection frequency $p_j^*$. The Invisible Fence selects the model
$M_j^*$ with the highest bootstrap selection frequency $p_j^*$. Jiang, Nguyen and Rao (\citeyear{Jiang2011})
apply the procedure to a genetic problem (which is not a
linear mixed model problem) with what they call a subtractive loss
$Q_M$ and show that, in this case, the Invisible Fence is very fast.
However, in general, including for linear mixed models, it is still
computationally burdensome to find the reduced set of models.

The idea of applying the Fence principle to subsets of the model space
rather than to the entire space to reduce the computation is developed
further in the Restricted Fence (\cite{Ng2012}). The basic
idea is to partition the model space $\mathcal{M}$ into not necessarily
disjoint subsets $\mathcal{M}_1, \ldots, \mathcal{M}_J$ and apply the
Simplified Adaptive Fence to each subset~$\mathcal{M}_j$. The final
model is then selected by applying the Simplified Adaptive Fence again
to select one of the $J$ already selected models. In particular
applications, the choice of subsets of the model space may be based on
substantive considerations, but it will often involve some arbitrary
choices. So just as the order of rows and columns affects the Cholesky
decomposition of $\PS$ (see Section~\ref{secnotation}) and hence can
affect model selection in shrinkage methods, the choice of subsets can,
in small samples, affect model selection with the Restricted Fence. The
Restricted Fence was introduced for selecting independent cluster
models when interest centers on the selection of the regression
parameters only and, in this case, the subsets $\mathcal{M}_j$
correspond to subsets of the columns\vadjust{\goodbreak} of $\XX$. It is then attractive to
multiply both sides of the model (\ref{eqremodel}) by a matrix that is
orthogonal to the columns of $\XX$ not in the current subset of
interest so that these variables are removed from the model. Two
further simplifications are introduced. First, instead of the
generalized least squares estimator (\ref{eqgls}) of the regression
parameters $\be$, \citet{Ng2012} use the least squares
estimator so that they do not have to estimate the marginal variance
matrix $\VV$. This involves a loss of efficiency but reduces the
convergence issues. Second, \citet{Ng2012} use a version of
the wild bootstrap in which they bootstrap from linear regression
models rather than linear mixed models. Both of these simplifications
are tied to selecting regression parameters, but they suggest useful
analogues for other problems and may be useful for model selection
methods beyond the Restricted Fence.
Generalizing and modifying the Restricted Fence to more general
situations is promising and deserves further attention.


\section{Other Bayesian Methods} \label{secBayes}

Bayesian model selection (also called model choice) requires us to
assign a prior distribution over $\mathcal{M}$ and compute the
posterior probabilities of each $M_l\in\mathcal{M}$. These
computations can be difficult so are usually carried out by applying
sophisticated Markov Chain Monte Carlo (MCMC) algorithms. We can
actually avoid explicit model selection by working directly with the
posterior distribution. If we need a single model, we can average over
$\mathcal{M}$ or we can select the model with highest posterior
probability. A useful way of interpreting this kind of selection (which
links it conceptually to shrinkage and Fence methods) is that the MCMC
algorithm reduces $\mathcal{M}$ to a small subset of models with
posterior probability above a threshold and we then select one of these.

For linear mixed models, this kind of approach has been explored in a
number of papers starting with \citet{Chen2003}. They consider
the problem of selecting the variance parameters $\ta$ in the
independent cluster model with $s_i=s_1$ and $\SI=\sigma^2\II_n$. They
introduce the alternative Cholesky decomposition and define $\bth^{\dagger}=(\be^T,\dd^{T},\ga^{\dagger T}, \sigma^2)^T$, where $\dd$
contains the diagonal elements of $\DD$ and $\ga^{\dagger}$ contains
the distinct elements of $\GA^{\dagger}$. Chen and Dunson assume that
the elements of $\dd$ are independently distributed with a point mass
at zero mixed with a Gaussian distribution truncated at zero. The
assumption that the elements of $\dd$ are independent allows each one
to be treated independently and the zero-inflated truncated-Gaussian
priors allow them to be exactly zero with positive probability.
Selection is based on running a Gibbs sampler and computing the
posterior probabilities of all possible models (of which there are at
most $2^{q_{\gamma}}$) by dividing the number of occurrences of each
model by the number of iterations. \citet{Saville09} point out
that these kinds of MCMC methods are generally time consuming to
implement, require special software and depend on subjective choice of
the hyperparameters in the priors.

As discussed in Section~\ref{secbic}, the problem can also be
formulated as a testing problem and the test carried out by computing
Bayes factors (\ref{eqbayesfac}). The two issues with using Bayes
factors are the choice of prior, which, depending on the formulation,
might need to include point mass at zero and should not be either too
concentrated or too dispersed, and the computation. \citet{Han2001} compare a number of methods for computing Bayes factors for
comparing two linear mixed models. They find that the reversible jump
(\cite{Green1995}) and marginal likelihood methods (\cite{Chib1995}) are able to
produce estimates of the Bayes factor and that the marginal likelihood
methods are easier to use. The marginal likelihood here refers not to
$\exp\{\ell(\bth)\}$ but, in the notation used to define the Bayes
factor (\ref{eqbayesfac}), to $\int g(\yy|\bth) h(\bth) \,d\bth$, where
$g(\yy|\bth) = \exp\{\ell(\bth)\}$ and $h$ is the prior for $\bth$.
\citet{Chib1995} and \citet{Han2001} comment that all the methods
require substantial human intervention and computer effort for a modest
payoff. These kinds of conclusions help motivate the use of
approximations like BIC (Section~\ref{secbic}) to the Bayes factor and
also more ad hoc alternative approaches to model selection such as
those of Spiegelhalter et al. (\citeyear{Spiegelhalter2002}) and Aitkin, Liu and Chadwick (\citeyear{Aikin09}).

Spiegelhalter et al. (\citeyear{Spiegelhalter2002}) propose a general Baye\-sian deviance
criterion for model selection of the form
\[
\DIC = \E\bigl\{-2\ell(\bth)|\yy\bigr\} + 2 \log\bigl\{f(\yy)\bigr\} +
2p_D,
\]
where $p_D = \E\{-\ell(\bth)|\yy\} + \ell(\bar{\bth})$, $\bar{\bth} = \E
(\bth|\yy)$ is the posterior mean of the parameters and $f(\cdot)$ is a
``fully specified standardizing term that is a function of the data
alone.'' The choice of $f$ is vague, but a natural choice is $f(\yy)=
\exp\{\ell(\hat{\bth})\}$ for some estimator $\hat{\bth}$ of $\bth$. If
the estimate $\hat{\bth}$ is fixed for all comparisons, then we can
omit the standardizing term. This is the same as just setting $f(\yy
)\equiv1$. For selecting regression terms in the mixed model when the
variance parameters $\ta$ are known, Spiegelhalter et al. (\citeyear{Spiegelhalter2002}) point
out that $p_D$ reduces to the effective degrees of freedom $\rho(\ta)$
defined in (\ref{eqedf}) so, as noted by \citet{Vaida2005},
DIC in this case is equivalent to marginal AIC with the asymptotic form
of the Vaida--Blanchard penalty for conditional AIC.

Aitkin, Liu and Chadwick (\citeyear{Aikin09}) propose a different way of using deviances to
select models from Spiegelhalter et al. (\citeyear{Spiegelhalter2002}). They suggest comparing
models $M_0$ and $M_1$ by computing the posterior distributions of the
parameters $\bth_{M_0}$ and $\bth_{M_1}$, generating $B$ realizations
$\bth_{M_0b}$ and $\bth_{M_1b}$ from the respective posterior
distributions and computing the empirical probability $\Pr\{-2\ell(\bth_{M_0b}) + 2\ell(\bth_{M_1b}) <\break -4.4|\yy\}$. The value $-4.4 =-2\log
(9)$ corresponds to a likelihood ratio of $9$ so the event $\{-2\ell
(\bth_{M_0b}) + 2\ell(\bth_{M_1b}) < -4.4\}$ represents strong evidence
for $M_0$ over $M_1$. They argue that if the empirical probability of
the event is $0.9$ or greater, there is a high posterior probability of
strong evidence in favor of $M_0$ over $M_1$. This approach has
attracted criticism from some Bayesians (Gelman, Robert and Rousseau,\break \citeyear{Gelman2010}).


\section{Simulation}\label{secsimulation}
Various authors have carried out simulations to compare different
methods of model selection, usually with one or more similar methods
and usually in problems with a small number of parameters. We review
some of these simulations in this section to see what we can learn from
putting the results together. Each simulation is limited but, together,
they are quite informative, particularly in identifying individual
problems in which particular methods work well. We think of this as
like a meta-analysis which extracts more information by combining
existing studies without having to repeat studies or run new studies. A
summary of the settings considered is given in Table~\ref{tblsimsum},
which is followed by a concise overview of the most important findings.
More detailed information and further comments on the simulations can
be found in the online supplementary material (see Appendix following
the bibliography).


%
\begin{table*}[t]
\caption{Table summarizing the
settings used in selected simulations. ``Reference'' shows the first
letters of the surnames of the authors and the last two digits of the
year of publication, ``Model'' describes the model considered, $m$ the
number of clusters and $n_i$ the size of the clusters. The quantities
$p$, $s_i$ and $q$ are the dimension of $\be$, the number of random
effects per cluster and the dimension of $\ta$ in the true model;
$p_f$, $s_{fi}$ and $q_f$ are the analogous quantities under the full
model. The next three measures describe the difficulty of selecting the
true model: $|\mathcal{M}_{\be}|$ and $|\mathcal{M}_{\ta}|$ are the
number of candidate models considered for $\be$ and $\ta$, respectively,
$\min|\beta_k|/\sigma$ measures the difficulty of selecting the
smallest nonzero regression parameter when there are no random effects
in the model and $\min\{$ev$(\PS/\sigma^2)\}$, the smallest eigenvalue
of $\PS/\sigma^2$, measures the difficulty of selecting the smallest
nonzero variance parameter. Finally, $\uu$ and $\ee$ describe the
distributions used for these random variables and ``Method'' denotes
the main model selection methods considered in the simulation}\label{tblsimsum}
\begin{tabular*}{\textwidth}{@{\extracolsep{\fill}}lccccc@{}}
\hline
\textbf{Reference} & \textbf{Model}& $\bolds{m/n_i}$ & $\bolds{p/p_f}$ & $\bolds{s_i/s_{fi}}$ & \multicolumn{1}{c@{}}{$\bolds{q/q_f}$} \\
\hline
CD03 & int${}+{}$slope & $200/8$ & $4/4$ & $3/4$ & $7/11$ \\[6pt]
DMT11& int${}+{}$slope & $10/\{6,26,51\}$ & $2/6$ & $2/2$ & $4/7$ \\
PN06 & int${}+{}$slope & $10/20$ & $3/5$ & $2/3$ & $3/7$ \\
SC08 & int & $\{15,20,30,50\}/3$ & $7/12$ & $1/1$ & $2/2$ \\
SC08 & int & $\{15,20,30,50\}/3$ & $4/5$ & $1/1$ & $2/2$ \\
GK10 & int & $\{10,20,40,80\}/\{3,6,9,12\}$ & $2/2$ & $1/1$ & $2/2$ \\[6pt]
DMT11 & int${}+{}$slope & $\{10,20,50\}/4$ & $2/6$ & $2/3$ & $4/7$ \\
DMT11 & int${}+{}$slope & $\{10,20,50\}/4$ & $3/6$ & $1/3$ & $4/7$ \\
SK10 & cluster & $20/\{1+\mathcal{B}(8,1/2)\}$ & $\{2,4,6\}/7$ & $\{1,2,3\}
/\{1,2,3\}$&$ 2/2$ \\
K11 & Fay--Herriot & $\{5,10, 30\}/1$ & $4/7$ & $1/1$ & $1/1$ \\
K11 & int & $\{5,10, 30\}/4$ & $\{2,4,6\}/\{5,7\}$ & $1/1$ & $2/2$ \\
JR03 & var comp & $8000/3$ & $2/5$ & $\{20,40\}/140$ & $\{2,3\}/8$ \\[6pt]
BKG10 & cluster & $\{30, 60\}/\{5, 10\} $& $2/9$ & $3/\{4,10\}$ & $7/\{11,
56\}$ \\
IZGG11 & cluster & $\{50, 100, 200\}/12$ & $3/8$ & $3/8$ & $7/37$ \\
PL12 & cluster &$\{10,20\}/\{10, 20\}$ & $3/5$ & $2/4$
& $3/10$ \\[6pt]
JRGN08 & Fay--Herriot & $30/1$ & $1\mbox{--}5/5$ & $1/1$ & $1/1$ \\
JRGN08 & int & $100/5$ & $\{2,4,5\}/5$ & $1/1$ & $2/$2 \\
JNR09 & int & $\{10,15\}/\mathcal{P}(3)$ & $\{3,6\}/6$ & $1/1$ & $2/2$ \\
NJ12 & int & $\{50, 100, 150\}/3$ & $7/30$ & $1/1$
& $2/2$ \\[9pt]
\hline
\textbf{Reference} & $\bolds{|\mathcal{M}_{\be}|/|\mathcal{M}_{\ta}|}$ & $\bolds{\min|\beta_k|/\sigma}$ & $\bolds{\min\{\mathrm{ev}(\PS/\sigma^2)\}}$ &
$\bolds{\uu/\ee}$ & \textbf{Method}\\
\hline
CD03& $1/16$ & 1 & 0.45 & $\mathcal{N}$ & Post prob.\\[6pt]
DMT11 & $14/3$ & 0.35 & 0.01 & $\mathcal{N}$ & IC \\
PN06 & $31/7$ & 0.2 & 0.5 & $\mathcal{N}$ & GIC\\
SC08 & $12/2$ & 1 & 2 & $\mathcal{N}$ & m{AIC} \\
SC08 & $31/2$ & 1 & 2 & $\mathcal{N}$ & m{AIC}\\
GK10 & $1/2$ & 1 & $\{0.1\mbox{--}0.8\}$ & $\mathcal{N}$ & cAIC \\[6pt]
DMT11 & $14/3$ & 1.83 & 0.17 & $\mathcal{N}/\{\mathcal{N}, \mathrm{mixtures}\}$ &
IC \\
DMT11 & $14/3$ & 1.83 & 0.11 & $\mathcal{N}/\{\mathcal{N}, \mathrm{mixtures}\}$ &
IC \\
SK10 & $7/1$ & 2 & $\{0.01,0.5,1\}$ & $\mathcal{N}$ & AIC \\
K11 & $7/1$ & 2 & 1 & $\{\mathcal{N}, \mathrm{mixture}(\mathcal{N},\mathcal{C})\}
$ &AIC \\
K11 & $7/2$ & 2 & $\{0.1, 1\}$ & $\{\mathcal{N}, t_3\}$ &AIC \\
JR03 & $31/\mathrm{NA}$ & 1.63 & 0.67 & $\mathcal{N}$ &own \\[6pt]
BKG10 & $512/16$ & 1 & 0.45 & $\mathcal{N}$ &shrinkage\\
IZGG11 & $256/256$ & $\{1.5,0.5\}$ & $\{0.41, 0.05\}$ & $\mathcal{N}$
&shrinkage \\
PL12 & $16/16$ & 1 & 0.32 &$\mathcal{N}$ & shrinkage \\[6pt]
JRGN08 & $32/1$ & 1 & 1 & $\mathcal{N}$ & Fence \\
JRGN08 & $32/1$ & 1 & 1 & $\mathcal{N}$ & AFence \\
JNR09 & $64/1$ & 1 & 1 & $\mathcal{N}$ & SAFence \\
NJ12 & $768/1$ & 0.001 & 1 & $\mathcal{N}$ & RFence \\
\hline
\end{tabular*}
\end{table*}

It is clear from Table~\ref{tblsimsum} that only a limited set of
models and limited settings have been considered. All except Srivastava
and Kubokawa (\citeyear{Srivastava2010}) and Jiang, Nguyen and Rao (\citeyear{Jiang2009}) considered the easier case
with constant\vadjust{\goodbreak} cluster size. The numbers of parameters and random
effects are very small in both the true and the full models; the
exceptions are Bondell, \mbox{Krishna} and
Ghosh (\citeyear{Bondell2010}) and \citet{Ibrahim2011} who
consider slightly larger numbers of variance parameters in the full
model ($q_f$) and \citet{Jiang2003} who consider large numbers of
random effects in the full model. The sets of candidate models are
relatively small, the largest occurring in Bondell, Krishna and
Ghosh (\citeyear{Bondell2010}) and
\citet{Ibrahim2011}. Small values of $\min|\beta_k|/\sigma$ and $\min
\{$ev$(\PS/\sigma^2)\}$ indicate that it is difficult to select the
true model for $\be$ and $\ta$, respectively. The table shows that, with
the exception of \citet{Ng2012}, the settings make it
relatively easy to select the true $\be$ and, surprisingly, often much
easier than to select the true $\ta$. This helps explain the general
conclusion that selecting $\be$ is easier than~$\ta$. Most authors
choose the true regression parameters according to their favored
procedure, that is, for AIC-like criteria $p$ is close to $p_f$ and
for BIC-like criteria and shrinkage methods $p$ is small compared to
$p_f$. Also, some authors apply their own variants of information
criteria without any justification or explanation, and possibly with
unintended effects.

For the marginal information criteria, as in linear regression models,
larger penalties tend to select smaller models, while smaller penalties
tend to select larger models. The bootstrap penalty is plausible
($m\mathit{AIC}_{B2}$ worked better than $m\mathit{AIC}_{B1}$) but has not been
thoroughly explored. For the conditional AIC criteria, the Greven--Kneib
penalty and the Srivastava--Kubokawa penalty produced promising results
but need a more thorough investigation. The philosophical differences
between using marginal and conditional criteria were explained by \citet{Vaida2005},
but the practical differences are much less
clear. Dimova, Markatou and
Talal (\citeyear{Dimova2011}) found in their simulation that the
conditional criteria performed worst at selecting the correct model,
tending to prefer larger models. They recommended GIC with
$a_n=n^{1/2}$ but noted that it does not always get the random effects
right, particularly when they have small variance. On the other hand, a
version of REML-based $m\mathit{AIC}_R$, which ignores the estimation of $\PS$,
worked well when $\PS$ is close to zero. Bondell, Krishna and
Ghosh (\citeyear{Bondell2010}) and
\citet{Ibrahim2011} obtained promising results for their shrinkage
methods. The methods of \citet{Ibrahim2011} have the advantage of
having two tuning parameters, although this makes the computations more
burdensome. They found that the SCAD penalty performed best for
regression parameters\vadjust{\goodbreak} and\break ALASSO for variance parameters. The Fence
methods can be difficult to implement with redundant variance
parameters and have not yet been investigated in the full model
selection problem. \citet{Chen2003} found that their approach
selected the true model with high probability and the performance was
robust to the choice of hyperparameters for the point mass at zero
mixed with a zero-truncated Gaussian distribution prior for each~$d_k$.

Finally, most of the studies used Gaussian distributions and those that
did not found that their methods performed more poorly under the
longer-tailed distributions they used.

\section{Discussion and Conclusions} \label{secdiscussion}

In this paper we have arranged, structured and reviewed a substantial
body of literature on different model selection procedures for linear
mixed models.
A key step in achieving this is our use of a unified notation for the
linear mixed model (\ref{eqremodel}), which we use in particular to
(i) bring together special cases of the linear mixed model such as the
variance component model, the independent cluster model, the clustered
variance component model, the random intercept and slope model, the
Fay--Herriot model or the longitudinal autoregression model; (ii) avoid
ambiguity in identifying what components are subject to selection:
regression parameters $\be$, variance parameters $\ga$, $\de$ or $\ta
=(\ga^T,\de^T)^T$ or both simultaneously, that is, $\bth= (\be^T,\ta^T)^T$; and (iii) make different model selection procedures suggested
by different authors easier to compare.

The performance of model selection procedures depends on how
performance is measured. Much of the theoretical work on information
criteria gives the right answer to a good question, such as how to
estimate the Akaike Information unbiasedly (AIC) or how to approximate
the Bayes factor accurately (BIC), but these criteria are not directly
related to model selection. Direct performance measures, such as how
often the data generating model or other correct models are detected,
are more useful. Parsimony (choosing models with few parameters) is an
important consideration when $p+q$ is large. It can be achieved by the
choice of combinations of the measure of model complexity, the penalty
function or the tuning constants and should be built into the
performance measures. Procedures that are optimal under one performance
measure need not be optimal under a different measure, so it may be
worthwhile to consider several measures.

One of the key issues in model selection is that the set $\mathcal M$
of candidate linear mixed models can be very large; depending on the
model, $\mathcal M$ can contain all $2^{p+q}$ possible models and, in
such cases, is very large when $p+q$ is large. Large candidate sets
$\mathcal M$ are computationally too demanding for methods like the
information criteria (Section~\ref{secinformation}) which try to
compare all the models in $\mathcal M$. A natural alternative approach
is to try to reduce $\mathcal M$ efficiently to a smaller subset of
models and then select models from within this subset. Shrinkage
methods (Section~\ref{seclasso}), Fence methods (Section \ref
{secfence}) and implicitly some Bayesian methods (Section \ref
{secBayes}) which try to do this are better able to handle large
$\mathcal M$. There are many open questions about how to reduce
$\mathcal{M}$ in appropriate ways and we anticipate an explosion of
results similar to that currently occurring in $n \ll p$ problems in
linear regression.

The theoretical treatment of mixed model selectors is difficult and
technical so the results that have been obtained are impressive.
Generally, these results require either strong assumptions or
restrictions to specific mixed models only, such as those having a
single variance parameter, and more theoretical insight would be very useful.

The difficulty of developing theoretical results\break means that we have to
rely on simulations to compare different methods. In reviewing the
various simulations, we found that only a limited set of models and
limited settings have been considered. In particular, the shrinkage and
Fence methods have only been applied to the independent cluster model
to date. As with the theory, more general and more challenging
scenarios should be investigated in the future. Interesting avenues for
future studies are to consider more general $\SI$ than $\sigma^2 \II_n$, letting $n$, $p$, $q$ and $s$ grow in different ways in asymptotic
studies, and exploring true joint selection of $\be$ and $\ta$.

With currently available software (e.g., \texttt{lmer} in R or \texttt
{Proc Mixed} in SAS), it is easy to initiate a request for and,
provided the problem is not too large or too sparse, to obtain a point
estimate for~$\bth$. Nonetheless, there are computational issues,
particularly when one or more variance parameters is zero (see Sections
\ref{seclasso} and~\ref{secfence}). This has implications for
computer intensive selection procedures, which can fail when estimation
in any one of the iterations fails. We expect that optimization
routines will develop and include better methods for dealing with
problems where the underlying model parameters are at or near the boundary.\vadjust{\goodbreak}
Similar and possibly more serious computational difficulties arise with
Bayesian methods. \citet{Han2001} remarked that all the Bayesian
methods they considered required substantial time and effort (both
human and computer). They pointed out that both the boundary issues and
the choice of priors have to be treated with care. 

There are interesting relationships between the method of estimation,
the method of selection and the definition of the possible model set
$\mathcal{M}$.
With clustered data, it is important to distinguish (\cite{Vaida2005}) or to be conscious of the distinction
(\cite{Greven2010}) between margin\-al questions regarding the underlying
population\break from which clusters are observed and conditional questions
regarding the particular clusters in the data when using information
criteria (Section~\ref{secinformation}). This distinction has
implications for shrinkage and Fence methods. Specifically, in order to
select models to treat conditional questions, it is worthwhile
developing shrinkage methods based on the conditional log-likelihood
$\ell(\bth|\hat\uu)$ and measuring model complexity in Fence using one
of the conditional AIC penalties in Table~\ref{tblcAICpenalties}.


\begin{appendix}
\section*{Appendix: Simulation Settings} \label{secAppendixsimulation}


\citet{Vaida2005} used as the full model the simple random
intercept and slope model
\begin{eqnarray}
y_{ij} = \beta_1 + x_{j}\beta_2
+ \zz^T_{ij}\GA_i\uu_i + \sigma
e_{ij},\nonumber \\
\eqntext{j = 1,\ldots, n_i\in\{6,26,51\}, i=1,\ldots, 10,}
\end{eqnarray}
with $\uu_i=(u_{1i},u_{2i})^T$, $\GA_i$ a $2\times2$ matrix of
parameters and $\zz^T_{ij} = (1, x_{j})$. The values\vspace*{1pt} of $x_j$ were
equally spaced in units of $5$ from $0$ to $25$ ($n_i=6$), $0$ to $125$
($n_i=26$) or $0$ to $250$ ($n_i=51$). The true models had $\be
=(-2.78, -0.186)^T$,
\[
\GA_i\GA_i^T = \PS_i =
\pmatrix{ 0.0367 & -0.00126
\vspace*{2pt}\cr
-0.00126 & 0.00279}
\]
and $\sigma^2 \in\{0.0705, 0.141, 0.282\}$.

\citet{Chen2003} reported results from a simulation using the
random intercept and slope regression model. In the part of the
simulation where they considered selecting $\bth$, the full model was
\begin{eqnarray*}
y_{ij}& =& \beta_1 + x_{2ij}\beta_2
+ x_{3ij}\beta_3 + x_{4ij}\beta_4 +
\zz^T_{ij}\DD_i \GA_i^{\dagger}
\uu_i \\
&&{}+ \sigma e_{ij},\quad  j = 1,\ldots, 8, i=1,\ldots, 200,
\end{eqnarray*}
with $\uu_i = (u_{1i}, u_{2i}, u_{3i}, u_{4i})^T$, $\DD_i$ a $4\times
4$ diagonal matrix, $\GA_i^{\dagger}$ a $4\times4$ matrix and $\zz^T_{ij} = (1, x_{2ij}, x_{3ij}, x_{4ij})$. The explanatory variables
were generated independently from the $\mathcal{U}(-2, 2)$
distribution.\vadjust{\goodbreak} The true mod\-el had $\be= \one_4$, $\DD_i = \diag(3, 1.2,
0.8, 0)$,
\[
\GA_i^{\dagger} = \pmatrix{ 1 &
0 & 0 & 0
\vspace*{1pt}\cr
1.33 & 1 & 0 & 0
\vspace*{1pt}\cr
0.25 & 0.71 & 1 & 0
\vspace*{1pt}\cr
0 & 0 & 0 & 0 }
\]
and $\sigma^2=1$. The set $\mathcal{M}$ of candidate models consisted
of all $2^4=16$ possible subsets of $\{d_1,\ldots, d_4\}$. \citet{Chen2003} used
a $\mathcal{N}_4(\zer_4,1000\II_4)$ prior for $\be$,
a Gamma $\mathcal{G}(0.05, 0.05)$ prior for $\sigma^{-2}$, a $\pi_0$
mixture of a point mass at zero and a $\mathcal{N}(0, 30)$ distribution
truncated at zero for each $d_k$ with $\pi_0 \in\{0.2, 0.5, 0.8\}$,
and independent $\mathcal{N}(0,0.5)$ distributions for the elements of
$\GA_i^{\dagger}$, given that they are nonzero.

\citet{Pu2006} carried out a simulation for the random intercept and
slope model
\begin{eqnarray*}
y_{ij} &=& \beta_1 + x_{2ij}\beta_2
+ x_{3ij}\beta_3 + x_{4ij}\beta_4 \\
&&{}+
x_{5ij}\beta_5 + \zz^T_{ij}
\GA_i\uu_i \\
&&{}+ \sigma e_{ij},\quad j = 1,\ldots, 20,
i=1,\ldots, 10,
\end{eqnarray*}
with $\uu_i=(u_{1i},u_{2i}, u_{3i})^T$, $\GA_i$ a $3\times3$ matrix of
parameters and $\zz^T_{ij} = (1, x_{2ij}, x_{3ij})$. The explanatory
variables were generated as independent $\mathcal{N}_4(\zer, \bA\bA^T)$
random vectors with
\[
\bA= \pmatrix{ 2.00 & 0.66 & 0.90 & 0.02
\vspace*{1pt}\cr
0.66 & 2.00 & 0.68 & 0.32
\vspace*{1pt}\cr
0.90 & 0.68 & 2.00 & 0.94
\vspace*{1pt}\cr
0.02 & 0.32 & 0.94 & 2.00 }.
\]
The true models had $\be= (\beta_1, 1.2, 0, 2.0, 0)^T$ with $\beta_1
\in\{0.5, 1.5, 0.2\}$, one of the variance matrices
\[
\GA_i\GA_i^T = \PS_i =
\pmatrix{ 1 & 0.5 & 0
\vspace*{1pt}\cr
0.5 & 1 & 0
\vspace*{1pt}\cr
0 & 0 & 0 }, \quad \pmatrix{1 & 0 & 0.5
\vspace*{1pt}\cr
0 & 0 & 0
\vspace*{1pt}\cr
0.5 & 0 & 1 }
\]
or
\[
\pmatrix{ 0 & 0 & 0
\vspace*{1pt}\cr
0 & 1 & 0.5
\vspace*{1pt}\cr
0 & 0.5 & 1 },
\]
and $\sigma^2=1$.
Following their suggested approach, \citet{Pu2006} included all
three random effects in the model and computed $\GIC$ with $a_n=\log(n)$
and $a_n=n^{1/2}$ for all $31$ candidate regression models. Then, using
the selected regression model, they computed the criteria over $7$
candidate variance models. They then iterated the process until the
selected models no longer changed.

\citet{Shang2008} reported a simulation using the random
intercept regression\vadjust{\goodbreak} model with $m \in\{15, 20, 30, 50\}$ and $n_i=3$
to compare the bootstrap AIC with $m\mathit{AIC}$. The full model included $12$
covariates (they do not explain how these were generated). The true
model had $p=7$ with $\be=\one_7$, $\gamma^2 = \Var(u_i) =2$ and $\sigma^2=1$.
The penalties were computed from $B=500$ parametric bootstrap
samples. Shang and Cavanaugh considered selecting the models with the
first covariate, the first two covariates, etc., and with or without
$u_{i}$. In a second simulation, they reduced the full model to $5$
covariates and for the true model set $p=4$ with $\be=\one_4$ and
considered all possible subsets of the $5$ variables and with or
without $u_{i}$.

\citet{Greven2010} carried out a simulation for penalized spline
smoothing and for the simple random intercept regression model
\begin{eqnarray}
y_{ij} = \beta_1 + x_{i}\beta_2
+ \gamma u_{i} +\sigma e_{ij},\nonumber\\
\eqntext{ j = 1,\ldots,
n_i\in\{3, 6, 9, 12\},} \\
\eqntext{i=1,\ldots, m\in\{10, 20, 40, 80\}.}
\end{eqnarray}
The clusters were taken to be of equal size in each run. The covariate
$x$ was chosen equally spaced in the interval $[0,1]$. The true models
had $\be= (0, 1)^T$, $\gamma^2 \in\{0, 0.1, 0.2, 0.4, 0.6, 0.8\}$ and
$\sigma^2 = 1$. The simulation compared the ability of $m\mathit{AIC}$, $c\mathit{AIC}$
with the asymptotic version of the Vaida--Blanchard penalty, the
Liang--Wu--Zhou (LWZ) penalty and the Greven--Kneib penalty to choose
between the simple linear regression model and the nonlinear or mixed
model. The nonlinearity in penalized spline smoothing is represented by
the random vector $\uu$, but there are only two variance parameters in
$\ta,$ so both the two models considered represent cases with a small
number of variance parameters.

In their simulation study, Dimova, Markatou and
Talal (\citeyear{Dimova2011}) compared a number of
different versions of marginal AIC ($m\mathit{AIC}$ with both finite sample and
asymptotic penalties, $m\mathit{AIC}$ treating $\PS$ as known, the REML version
$m\mathit{AIC}_R$, $m\mathit{AIC}_R$ treating $\PS$ as known), conditional AIC ($c\mathit{AIC}$
with both finite sample and asymptotic penalties and $c\mathit{AIC}$ using the
REML estimates with both finite sample and asymptotic penalties), BIC
[which is GIC with $a_n=\log(n)$] and GIC with $a_n=n^{1/2}$. The full
model was the random intercept and slope model
\begin{eqnarray}
y_{ij} &=& \beta_1 + x_{2i}\beta_2
+ x_{3i}\beta_3 + x_{4i}\beta_4 +
x_{5ij}\beta_5\nonumber\\
&&{} + x_{5ij}^2
\beta_6 + \zz^T_{ij}\GA_i
\uu_i\nonumber\\
&&{} + \sigma e_{ij},\quad j=1,\ldots, 4,\nonumber\\
 \eqntext{i=1,\ldots,m \in\{10, 20, 50\},}
\end{eqnarray}
with $\uu_i=(u_{1i},u_{2i}, u_{3i})^T$, $\GA_i$ a $3\times3$ matrix of
parameters and $\zz^T_{ij}=(1, x_{5ij}, x_{5ij}^2)$. The explanatory
variables $x_{2i} \sim \mathrm{independent}\ \mathcal{N}(0,1)$, $x_{3i}$
and $x_{4i}$ were generated from the $\mathcal{N}(3,4)$ distribution,
and $x_{5i1} =0$, $x_{5i2}=6$, $x_{5i3}=12$ and $x_{5i4}=24$ so $\xx_{5i} = (0, 6,\break 12, 24)^T$. The $u_{1i}$'s were generated from Gaussian
distributions, the $e_{ij}$ were generated from Gaussian or Gaussian
mixture distributions $\zeta\mathcal{N}(0, 1.2) + (1-\zeta) \mathcal
{N}(8, 16)$ with $\zeta\in\{0.9, 0.8, 0.6\}$.
The two true models considered had (a) $\be= (3,2,0,0,0,0)^T$, the
$(1,1)$ entry $\psi$ of $\PS_i = \GA_i\GA_i^T$ satisfying $\psi=\Var
(u_{1i}) \in\{0.2,\break 0.5, 1.5, 4\}$ with all other entries zero, and
$\sigma^2 = 1.2$, and (b) $\be=(10,5,0,0,2,0)$,
\[
\GA_i\GA_i^T = \PS_i =
\pmatrix{ 4 & 0.5 & 0
\vspace*{2pt}\cr
0.5 & \psi& 0
\vspace*{2pt}\cr
0 & 0 & 0 },
\]
with $\psi\in\{0.2, 0.5, 1.5, 4\}$, and $\sigma^2 = 1.2$. Dimova et~al. fitted $42$
candidate models to the data. These includ\-ed $6$ models
with $u_{1i}$, with $u_{1i}$ and $x_{4ij}$, with $u_{1i}$ and
$(x_{4ij},x_{4ij}^2)$, with $(u_{1i}, u_{2j})$~and $x_{4ij}$, with
$(u_{1i}, u_{2j})$ and $(x_{4ij}, x_{4ij}^2)$, and with
$(u_{1i},u_{2i}, u_{3i})$ and $(x_{4ij}, x_{4ij}^2)$, crossed with
models for the regression structure made up of the $2^3-1=7$ subsets of
$(x_{1i}, x_{2i},x_{3i})$.

\renewcommand{\thetable}{\arabic{table}}
\setcounter{table}{2}
\begin{table*}
\def\arraystretch{0.9}
\caption{Simulation settings for the simulation reported by Kubokawa (\citeyear{Kubokawa2011}). The first two cases are from the Fay--Herriot model in which
$n_i=1$ so $n=m$; the last eight are from the random intercept
regression model with $n_i=4$ so $n=4m$. In \textup{I}-1 and \textup{I}-2, $\sigma^2$ is
treated as known. \textup{II}-1 and \textup{II}-2 use the same settings, but in \textup{II}-1 the
variance parameters are treated as known. Also, $p_f$ is the dimension
of the regression parameter $\be$ in the largest candidate model and
$p$ is the dimension of the regression parameter in the true model. In
the mixture models, $\mathcal{C}$ denotes the Cauchy distribution}
\label{tbltab2}
\begin{tabular*}{\textwidth}{@{\extracolsep{\fill}}lccd{2.0}d{1.2}d{1.2}cc@{}}
\hline
\textbf{Code} & $\bolds{p_f}$ & \multicolumn{1}{c}{$\bolds{p}$} & \multicolumn{1}{c}{$\bolds{m}$} & \multicolumn{1}{c}{$\bolds{\gamma^2}$} & \multicolumn{1}{c}{$\bolds{\sigma^2}$} & $\bolds{\uu}$ & $\bolds{\ee}$ \\
\hline
I-1 & 7 & 4 & 10 & 0.25 & 0.25 & $\mathcal{N}(0,1)$ & $\mathcal
{N}(0,1)$ \\
I-2 & 7 & 4 & 50 & 0.25 & 0.25 & $0.9\mathcal{N}(0,1) + 0.1 \mathcal
{C}$ & $0.9\mathcal{N}(0,1) + 0.1 \mathcal{C}$\\[6pt]
II-1 & 7 & 4 & 10 & 0.1 & 1 & $\mathcal{N}(0,1)$ & $\mathcal{N}(0,1)$ \\
II-2 & 7 & 4 & 10 & 0.1 & 1 & $\mathcal{N}(0,1)$ & $\mathcal{N}(0,1)$ \\
II-3 & 7 & 6 & 5 & 1 & 1 & $t_3$ & $t_3$ \\
II-4 & 7 & 2 & 30 & 1 & 1 & $t_3$ & $\mathcal{N}(0,1)$ \\
III-1 & 5 & 2 & 5 & 0 & 1 & -- & $\mathcal{N}(0,1)$ \\
III-2 & 5 & 4 & 30 & 0 & 1 & -- & $t_3$\\
III-3 & 5 & 2 & 5 & 1 & 1 & $\mathcal{N}(0,1)$ & $\mathcal{N}(0,1)$\\
III-4 & 5 & 4 & 30 & 1 & 1 & $t_3$ & $t_3$\\
\hline
\end{tabular*}   \vspace*{-3pt}
\end{table*}

\begin{table*}[b]\vspace*{-3pt}
\def\arraystretch{0.9}
\caption{The $n \times20$ matrices $\ZZ^{(j)}$ used by Jiang and Rao (\citeyear{Jiang2003}) in their simulation. Here $\II_m$ is the $m\times m$ identity
matrix, $\one_m$ is the $m$-vector of ones and $\otimes$ is the
Kronecker product}\label{tab3}
\begin{tabular*}{\textwidth}{@{\extracolsep{\fill}}lcc@{}}
\hline
$\bolds{L_3}$ & $\bolds{L_2}$ & $\bolds{L_1}$\\
\hline
$\ZZ^{(1)} = \II_{20} \otimes\one_{20} \otimes\one_{20} \otimes\one_{3}$ & $\ZZ^{(4)} = \II_{20} \otimes\II_{20} \otimes\one_{20}
\otimes\one_{3}$ & $\ZZ^{(7)} = \II_{20} \otimes\II_{20} \otimes\II_{20} \otimes\one_{3}$\\
$\ZZ^{(2)} = \one_{20} \otimes\II_{20} \otimes\one_{20} \otimes\one_{3}$ & $\ZZ^{(5)} = \II_{20} \otimes\one_{20} \otimes\II_{20}
\otimes\one_{3}$ &\\
$\ZZ^{(3)} = \one_{20} \otimes\one_{20} \otimes\II_{20} \otimes\one_{3}$ & $\ZZ^{(6)} = \one_{20} \otimes\II_{20} \otimes\II_{20}
\otimes\one_{3}$ & \\
\hline
\end{tabular*}
\end{table*}

Srivastava and Kubokawa (\citeyear{Srivastava2010}) carried out a simulation study using the
independent cluster mod\-el (\ref{eqremodelg}) with $m=20$ clusters of
size $n_i \sim1 + \mathcal{B}(8, 1/2)$, where $\mathcal{B}$ denotes
the binomial distribution. The full model had~$7$ explanatory variables
with $s_i \equiv s_1 \in\{1,2,3\}$ random effects in each cluster. The
$n_i$ rows of $\XX_i$ were generated independently from the $\mathcal
{N}_7(\zer_7,\break 0.7\II_7 + 0.3\JJ_7)$ and the $n_i$ rows of $\ZZ_i$ were
generated independently from $\mathcal{N}_{s_1}(\zer_{s_1}, 0.7\II_{s_1} + 0.3\JJ_{s_1})$. The true models had $p \in\{2,4,6\}$
explanatory variables and the same random effects as the full model, as
only selection on the regression parameters was considered. Srivastava
and Kubokawa set $\beta_k =\break 2(-1)^{k+1}\{1+\mathcal{U}(0,1)\}$, for $k
= 1,\ldots, p$, $\PS=\gamma^2\II_s$ with $\gamma^2 \in\{0.01, 0.5, 1\}
$ and $\sigma^2=1$. The $7$ candidate models had the correct variance
structure and the first, first two, first three explanatory variables,
etc. The simulation consisted of $10$ generated values of $\XX$ and $\ZZ
$ with $30$ sets of $\yy$ for each, making $300$ replications. They
reported the frequency of selecting the correct model for $(p=2, \psi
=0.01, s_1=1)$, $(p=4, \break\psi=0.5, s_1=2)$ and $(p=6, \psi=1, s_1=3)$ for
both known and unknown $\gamma^2$. The Srivastava--Kubokawa conditional
AIC method (\ref{eqSKcAIC}) using the different estimators of $\be$
and $\vv$ performed similarly and outperformed $m\mathit{AIC}$ and $c\mathit{AIC}$ which
were also very similar. A second simulation carried out with $\PS=
\diag(\psi_1,\ldots, \psi_{s_1})$ produced similar results.\vadjust{\goodbreak}

\citet{Kubokawa2011} carried out simulations under the Fay--Herriot model and
the random intercept regression model, essentially comparing marginal
and conditional AIC criteria with his Mallows type criteria. The
Fay--Herriot model can be viewed as a special case of the random
intercept regression model in which $n_i=1$ so $n=m$ and $\Var(e)=\sigma^2$ is known; in the simulation, the random intercept regression model
had clusters of size $n_i=4$ so $n=4m$. The full models had $p_f \in\{
5, 7\}$ explanatory variables; the $n_i$ rows of $\XX_i$ were generated
independently from the $\mathcal{N}_{p_f}(\zer_{p_f}, 0.7\II_{p_f} +
0.3\JJ_{p_f})$ distribution. The components of $\uu$ and $\ee$ were
generated independently from various distributions. The true models had
$p \in\{2,4,6\}$ explanatory variables with the nonzero coefficients
$\beta_k = 2(-1)^{k+1}\{1+\mathcal{U}(0,1)\}$, $k = 1,\ldots, p$ and
various values of $\gamma^2$ and $\sigma^2$. A full list of settings is
given in Table~\ref{tbltab2}. The simulation was carried out by
generating $20$ values of $\XX$ and $50$ sets of $\yy$ for each value
of $\XX$, making $1000$ replications. In settings I and II, the $7$
candidate models had the correct variance structure so only selection
on the regression parameter including the first, first two, first three
regressors, etc.\vadjust{\goodbreak} was considered. In setting~III, the models were also
considered with and without the variance structure. The criteria all
performed similarly in the first two settings (although, as noted in
Section~\ref{secother}, the Mallows criteria estimating $\ta$ from the
candidate model performed very poorly) and $c\mathit{AIC}$ was superior for the
Fay--Herriot model but slightly inferior for the random intercept
regression model. For selection on all the parameters, $m\mathit{AIC}$ and
$c\mathit{AIC}$ worked well, but $m\PEC$ was poor and $c\PEC$ tended to select
models without random effects. Kubokawa concluded that these criteria
are only useful for selecting regression parameters.

\citet{Jiang2003} reported results from a simulation using a crossed
three factor regression model. In the part of the simulation where they
considered selecting $\bth$, the full model with $m_j=20$, $n_i=3$ (so
the sample size $n = 3\times20^3 = 24\mbox{,}000$) was
\[
\yy= \XX\be+ \sum_{j=1}^7
\gamma_j\ZZ^{(j)}\uu_j + \sigma\ee,
\]
with $\XX$ a $n\times5$ matrix, $\be$ a $5$-vector, $\ZZ^{(j)}$ the $n
\times20$ matrices defined in Table~\ref{tab3} and $\uu_j$ independent
random $20$-vectors.\vadjust{\goodbreak} The explanatory variables were generated as
standard Gaussian random variables. They do not explain how the $\uu_{j}$'s and $\ee$ were generated, but they are most likely standard
Gaussian. The two true models considered both had $\be= (2,0,0,4,0)^T$
and either $\ga= (1,\zer_6^T)^T$ or $\ga= (1,0, 1.5, \zer_4^T)^T$. In
both cases $\sigma^2 = 1.5$. Jiang and Rao did not specify the set of
candidate models; implicitly it is the set of $2^5-1=31$ all possible
regression models multiplied by the number of choices in each of $L_1$,
$L_2$ and $L_3$. However, $L_2$ and $L_1$ contain the two-way and
three-way interactions of the terms in $L_3$, so it would be more usual
to select from $L_1$ and, only if the model in $L_1$ is not selected,
select from $L_2$, allowing the results of this selection to determine
what we consider for selection in $L_3$. Jiang and Rao found that the
penalty $a_n = a_{jn} = n/\log(n)$ worked well.\looseness=-1

Bondell, Krishna and
Ghosh (\citeyear{Bondell2010}) undertook a simulation study to examine the
properties of their LASSO procedure in finite sample size settings. The
full model was the independent cluster model
\begin{eqnarray}
\yy_i = \XX_i\be+ \ZZ_i\GA_i
\uu_i + \sigma\ee_i, \quad j=1,\ldots, n_i \in\{
5,10\},\nonumber\\
\eqntext{ i=1,\ldots, m\in\{30,60\},}
\end{eqnarray}
with equal size clusters in each run, $\XX_i$ a $n_i \times9$ matrix
of independent $\mathcal{U}(-2,2)$ random variables, $\ZZ_i$ either a
$n_i \times4$ matrix independent of $\XX$ with first column $\one_{n_i}$ and the remaining $3$ columns generated from the $\mathcal
{U}(-2,-2)$ distribution [when $(n_i,m) \in\break\{(5,30), (10, 60)\}$] or
$\ZZ_i=(\mathbf{1}_{n_i}, \XX_i)$ a $n_i \times10$ matrix [when
$(n_i,m) =(5, 60)$], $\GA_i$ either a $4 \times4$ or a $10 \times10$
matrix, $\uu_i$ either a $4$- or a $10$-vector, $\sigma$ a scalar and
$\ee_i$ an $n_i$-vector. For the true model, for the first two
scenarios $(n_i,m) \in\{(5,30), (10, 60)\}$, they set $\be=(1,1, \zer_7^T)^T$ and for the third $(n_i,m) =(5, 60)$, they set $\be=(1,0,1,
\zer_6^T)^T$. In all three scenarios,
\[
\GA_i\GA_i^T =\PS_i=
\pmatrix{ 9 & 4.8 & 0.6 \vspace*{2pt}
\cr
4.8 & 4 & 1 \vspace*{2pt}
\cr
0.6 & 1 & 1 } %
\]
and $\sigma=1$. Bondell, Krishna and
Ghosh (\citeyear{Bondell2010}) compared their model selection
procedure with the earlier approach suggested in the literature which
first selects either the regression or variance structure using AIC
and/or BIC while fixing the other at the full model (e.g., \cite*{Pu2006}). As a further comparison they also applied the ALASSO, LASSO and
a stepwise procedure for selecting $\be$ given $\ta$ after first
selecting $\ta$ by fixing $\be$ at the full model. The new procedure
was shown to be closest to ``oracle'' and to correctly identify the true
model most often.

\citet{Ibrahim2011} also undertook a simulation study and considered\vadjust{\goodbreak}
six different scenarios for the independent cluster model. They
considered the model
\begin{eqnarray}
\yy_i = \XX_i\be+ \ZZ_i\GA_i
\uu_i + \sigma\ee_i,\quad j=1,\ldots, 12, \nonumber\\
\eqntext{i=1,\ldots, m\in
\{50,100, 200\},}
\end{eqnarray}
where $\XX_i$ is a $12 \times8$ matrix with independent rows $\xx_{ij}^T$ and $\xx_{ij} \sim\mathcal{N}_8(\zer_8, \SI_{X})$, $\SI_X=(0.5^{\mid r-s\mid})$, $\ZZ_i = \XX_i$, $\GA_i$ is a $8 \times8$
matrix, $\uu_i$ is a $8$-vector, $\sigma$ is a nonnegative scalar and
$\ee_i$ a $12$-vector. For the true model they set $\be=(3,2,1.5,\zer_5)^T$,
\begin{eqnarray}
\GA_i\GA_i^T =\PS_i=
\pmatrix{ \PS_i^* & \zer_{3\times5} \vspace*{2pt}
\cr
\zer_{5 \times3} & \zer_{5\times5} }\nonumber\\
\eqntext{\displaystyle\mbox{with }
\PS_i^* = %
\pmatrix{ 1 & 0.5 & 0.25 \vspace*{2pt}
\cr
0.5 &
1 & 0.5 \vspace*{2pt}
\cr
0.25 &0.5 & 1} %
,}
\end{eqnarray}
and $\sigma\in\{1,3\}$. The full model contains $5$ unnecessary sets
of random effects in each cluster. The simulation study concluded that
for selecting the regression parameters, the SCAD penalty performed
best in terms of estimation error and minimizing overfit. For the
variance parameters, the ALASSO penalty performed best. In all cases
the penalized maximum likelihood estimates performed better than the
maximum likelihood estimates from the full model.

\citet{Peng2012} carried out a simulation using the same setting as
Bondell, Krishna and
Ghosh (\citeyear{Bondell2010}) and then their own setting to examine the
properties of their shrinkage method in finite sample size settings.
The full model was the independent cluster model
\begin{eqnarray}
\yy_i = \XX_i\be+ \ZZ_i\GA_i
\uu_i + \sigma\ee_i,\quad j=1,\ldots, n_i \in\{
10,20\},\nonumber\\
\eqntext{ i=1,\ldots, m\in\{10,20\},}
\end{eqnarray}
with equal size clusters in each run, $\XX_i$ a $n_i \times5$ matrix
of independent standard Gaussian random variables, $\ZZ_i$ a $n_i
\times4$ matrix with columns equal to the first $4$ columns of $\XX_i$, $\GA_i$ a $4 \times4$ matrix, $\uu_i$ a $4$-vector, $\sigma$ a
scalar and $\ee_i$ an $n_i$-vector. For the true model, they set $\be
=(1,0,1.5, 1,0)^T$,
\[
\GA_i\GA_i^T =\PS_i=
\pmatrix{ 0 & 0 & 0 & 0 \vspace*{2pt}
\cr
0 & 0.5 & 0 & 0.354
\vspace*{2pt}
\cr
0 & 0 & 0 & 0 \vspace*{2pt}
\cr
0 & 0.354 & 0 & 1} %
\]
and $\sigma=1$. In their first simulation using the same setting as
Bondell, Krishna and
Ghosh (\citeyear{Bondell2010}), \citet{Peng2012} compared the effect of using
different methods to select the tuning parameters in their method.
Their conclusion is that BIC is the best method of selecting their
tuning parameters.\vadjust{\goodbreak} This conclusion is based on the average percentage
of coefficients that are incorrectly estimated to be non\-zero, the
average percentage of coefficients that are incorrectly estimated to be
zero, the average size of the selected model and the probability of
identifying the correct model. The first 3 measures are all marginal
measures which are less stringent criteria than the probability of
identifying the correct model. They use the simulation probability of
identifying the correct model to compare their results with the
reported
results of Bondell, Krishna and
Ghosh (\citeyear{Bondell2010}), that is, without recalculating these
estimates for their data. Their method performs very poorly for the
smaller sample sizes but well for the larger sample sizes. They used
the second simulation setting to compare their parameter estimates in
the selected model with the maximum likelihood estimators for the true
model and showed that their performance is comparable.

\citet{Jiang2008} illustrated the use of the Adaptive Fence method in
two scenarios. The first is the Fay--Herriot model and they showed that
if the data generating model is
\[
y_i = \xx^T_i\be+\gamma
u_i+e_i, \quad i=1,\ldots,30,
\]
where $u_i$, $e_i\sim\mathrm{independent}\ \mathcal{N}(0,1)$, then, for
$M_c\subset M_f$, the quantity $\sigma_{M_c,M_f}$ is completely known.
They considered only selection on $\be$ and did not compare their
results with other selection procedures. The true models had $\be^T =
(1,0,0,0,0),$ $(1,2,0,0,0)$, $(1,2,3,0,0)$, $(1,2,3,2,0)$,
$(1,2,3,2,3)$ and $\gamma=1$.\break \citet{Jiang2008} reported simulation
results based on $100$ runs for a range of choices of the tuning
parameter $b_n$ in equation (\ref{eqnfence1}).
The second scenario is the random intercept regression model
\begin{eqnarray}
y_{ij} = \xx^T_{ij}\be+ \gamma
u_i + \sigma e_{ij},\quad j = 1, \ldots,5 ,\nonumber\\
 \eqntext{i=1,\ldots ,100,}
\end{eqnarray}
where $\be$ is a $5$-vector and $\gamma$ and $\sigma$ are scalar. They
generated $u_i\sim \mathrm{independent}\ \mathcal{N}(0,1)$, $\ee_{i} =
(e_{i1},\ldots,\break  e_{i5})^T\sim \mathrm{independent}\ \mathcal{N}(\zer_5,(1-\zeta)\II_5 + \zeta\JJ_5)$, $\zeta\in\break\{0,0.2,0.5,0.8 \}$ and
$x_{ij2},\ldots,x_{ij5} \sim  \mathrm{independent}\break \mathcal{N}(0,1)$ so $\xx^T_{ij} = (1,x_{ij2},\ldots,x_{ij5})$. The true models had $\be^T =
(2,0,0,4,0)$, $(2,9,0,4,8)$, $(1,2,3,2,3)$ and $\gamma=\sigma=1$. As a
lack-of-fit measure they choose the residual sum of squares and showed
that the Adaptive Fence chooses the true model in all $100$ simulation
runs. In comparison, the performance of two GIC type criteria as
introduced in \citet{Jiang2003} is less impressive, particularly
when the true model is the full model and $\zeta$ is large.

Jiang, Nguyen and Rao (\citeyear{Jiang2009}) reported limited simulation results using the
Simplified Adaptive\break Fence for a different random intercept regression model, but again selection only focused on the regression parameters
$\be^T = (\beta_1,\ldots,\beta_6)$. The model was
\begin{eqnarray}
y_{ij} = \xx^T_{ij}\be+ \gamma
u_i + \sigma e_{ij}, \quad j = 1, \ldots,n_i \sim\mathcal{P}(3),\nonumber\\
 \eqntext{i=1,\ldots ,m\in\{10,15\},}
\end{eqnarray}
where $\gamma$ and $\sigma$ are scalar, $u_i,e_{ij},x_{ij1},x_{ij2}
\sim  \mathrm{indepen\mbox{-}}\break\mathrm{dent}\ \mathcal N(0,1)$\vspace*{1pt} and $\xx^T_{ij} =
(1,x_{ij1},x_{ij2},x_{ij1}^2,x_{ij2}^2, x_{ij1}x_{ij2})$. A total of
$100$ simulation runs were run with two true models with $\be^T =
(1,1,1,0,0,0)$ and $\be= \mathbf{1}_6$ (i.e., the full model) and
$\gamma=\sigma=1$. As a performance measure the number of times the
true model was selected was used and the reported results only showed
the selection probabilities, which seem to be good, without comparing
them to other selection criteria.

\citet{Ng2012} evaluated the Restricted Fence method in a
simulation based on data from a bone turnover study. The setting is the
random intercept regression model
\begin{eqnarray}
y_{ij} = \xx^T_{ij}\be+ \gamma
u_i + \sigma e_{ij},\nonumber\\
 \eqntext{j =
1,\ldots, 3, i=1,\ldots ,m=\{50, 100, 150\},}
\end{eqnarray}
where $\be$ is a $30$-vector and $\gamma$ and $\sigma$ are scalar. They
generated $u_i\sim \mathrm{independent}\ \mathcal{N}(0,1)$, $\ee_{i} =
(e_{i1},\ldots,\break e_{i3})^T\sim \mathrm{independent}\ \mathcal{N}(\zer_3,\II_3)$,
one explanatory variable corresponding to dietary group as binary
and the remaining explanatory variables as independent Gaussian
variables with means and variances the same as those for the variables
in the bone turnover study. The true models had $7$ variables in the
mean and $\gamma=\sigma=1$. For the Restricted Fence, the variables
were divided into $4$ subsets of $7$ or $8$ variables using biological
considerations and $100$ bootstrap samples used in each selection.
\citet{Ng2012} compared the Restricted Fence with particular
backward and forward search implementations of information criteria.
They showed that the Restricted Fence underfits when the sample size is
small but performs well when the sample size increases. They found that
the information criteria tend to overfit and that BIC performed best of
the information criteria.
\end{appendix}

\section*{Acknowledgments}

This research was supported by an Australian Research Council discovery project grant.
We thank two referees and an Associate Editor for
their reviews which have lead to an improved paper.



\end{document}